\documentclass[aps,pra,nobalancelastpage,superscriptaddress, 10pt, twocolumn]{revtex4-2}

\usepackage{color,ifthen,amsthm,amsmath,amsxtra,amsfonts,dsfont,graphicx,bm,tikz,scalerel,wasysym,bbm,graphicx,amsthm,braket,empheq,CircuitTikz,physics}
\usepackage[colorlinks=true,linkcolor=blue, citecolor=blue, urlcolor=blue, bookmarks]{hyperref}
\usepackage[capitalize]{cleveref}  

\usepackage{comment}
\usepackage{booktabs}

\newcommand{\be}{\begin{equation}}
\newcommand{\ee}{\end{equation}}
\newcommand{\bea}{\begin{eqnarray}}
\newcommand{\eea}{\end{eqnarray}}
\newcommand{\bse}{\begin{subequations}}
\newcommand{\ese}{\end{subequations}}

\theoremstyle{plain}
\newcommand{\1}{\mathbbm{1}}

\newcommand{\ruc}{\rm RUC}
\newcommand{\rpc}{\rm RPC}

\newtheorem{lemma}{Lemma}
\theoremstyle{plain}

\theoremstyle{plain}

\definecolor{mygreen}{RGB}{80, 150, 80}
\definecolor{deepgreen}{RGB}{000, 80, 000}

\usetikzlibrary{external}

\begin{document}

\author{Bruno Bertini}
\affiliation{School of Physics and Astronomy, University of Birmingham, Edgbaston, Birmingham, B15 2TT, UK}

\author{Katja Klobas}
\affiliation{School of Physics and Astronomy, University of Birmingham, Edgbaston, Birmingham, B15 2TT, UK}

\author{Pavel Kos}
\affiliation{Max-Planck-Institut f\"ur Quantenoptik, Hans-Kopfermann-Str. 1, 85748 Garching}

\author{Daniel Malz}
\affiliation{Department of Mathematical Sciences, University of Copenhagen, Universitetsparken 5, 2100 Copenhagen, Denmark}

\title{Quantum and Classical Dynamics with Random Permutation Circuits}

\begin{abstract}
  Understanding thermalisation in quantum many-body systems is among the most enduring problems in modern physics. A particularly interesting question concerns the role played by quantum mechanics in this process, i.e.\ whether thermalisation in quantum many-body systems is fundamentally different from that in classical many-body systems and, if so, which of its features are genuinely quantum. Here we study this question in minimally structured many-body systems that are only constrained to have local interactions, i.e.\ local random circuits. 
In particular, we introduce random permutation circuits (RPCs), which are circuits comprising gates that locally permute basis states, as a counterpart to random unitary circuits (RUCs), a standard toy model for generic quantum dynamics. RPCs represent a model for generic microscopic classical reversible dynamics, but, interestingly, can be interpreted both as classical or as quantum dynamics. 
  We show that, upon averaging over all circuit realizations, RPCs permit the analytical computation of several key quantities such as out-of-time order correlators (OTOCs) and entanglement entropies. In the classical setting, we obtain similar exact results relating (quantum) purity to (classical) growth of mutual information and (quantum) OTOCs to (classical) decorrelators. 
  We thus discover a series of exact relations, connecting quantities in RUC and (quantum or classical) RPCs.
Our results indicate that despite the fundamental differences between quantum and classical systems, their many-body dynamics exhibits remarkably similar behaviours.
\end{abstract}

\maketitle


\section{Introduction}
Finding universal aspects in the dynamics of many-body quantum systems --- and the emergent laws governing them --- is one of the central themes of modern theoretical physics. Although this question is as old as quantum mechanics~\cite{von2010proof}, the last two decades have brought important progress in our understanding. The key driving factors have been the advent of new approaches borrowed from other fields of physics --- such as quantum information~\cite{prosen2007operator, hayden2007black, sekino2008fast, hosur2016chaos, brandao2016efficient}, high energy physics~\cite{calabrese2005evolution, liu2014entanglement, shenker2014black, shenker2014multiple,maldacena2016bound}, and quantum chaos~\cite{chen2018universal, kos2018many, chan2018solution, bertini2018exact} --- and the discovery of new methods to characterise out-of-equilibrium quantum matter. Two important examples are those based on the combination of random matrix theory and spatial locality~\cite{nahum2017quantum,fisher2022random}, and those exploiting the duality between space and time~\cite{bertini2018exact,bertini2019exact,ippoliti2021postselection,bertini2022growth}. These approaches turned out to be especially powerful in systems, dubbed \emph{quantum circuits}, which are defined in a discrete space-time and led to a much deeper understanding of quantum dynamics. For instance, we now know how the support of local operators expands under the time evolution~\cite{nahum2018operator,chan2018solution,vonKeyserlingk2018operator,rakovszky2018diffusive, alba2019operator, bertini2020operatorI, bertini2019exact, bertini2020scrambling, claeys2020maximum, dowling2023scrambling} and how the entanglement between different spatial regions grows~\cite{nahum2017quantum,bertini2018entanglement, gopalakrishnan2019unitary, zhou2019emergent, zhou2020entanglement, vasseur2019entanglement, vonKeyserlingk2018operator, zhou2022maximal, foligno2023growth, richter2023transport}.  

A natural question arising from this recent success is what is precisely the role of ``quantumness'' in the emergence of this phenomenology. Could similar physics be observed in classical systems? The closest classical analogue of quantum circuits are reversible classical circuits (such as cellular automata and Boolean circuits), which have long been used to model universal features of classical dynamics~\cite{Aldana2003}. For instance, recent work has introduced objects known as \emph{decorrelators}~\cite{Das2018,Bilitewski2018}, which probe operator spreading in the classical setting much like out-of-time-order correlators (OTOC)s~\cite{larkin1969quasiclassical,shenker2014black,shenker2014multiple,maldacena2016bound} do in the quantum setting. Analogously to OTOCs, decorrelators can be used to diagnose the speed of operator spreading, the so called \emph{butterfly velocity}, and obtain Lyapunov exponents~\cite{Liu2021a}. Similarly, in reversible classical circuits the mutual information across a bipartition~\cite{cover1999elements} closely matches the behaviour of entanglement entropy in quantum chaotic dynamics. This includes both the linear growth of mutual information to a volume law and even exhibiting a form of measurement-induced phase transition~\cite{Pizzi2022,Pizzi2024}.

The local updates in reversible classical circuits are implemented by permutations, which are particular examples of unitary matrices, and therefore classical circuits can be interpreted as special quantum circuits. The corresponding quantum model exhibits the same dynamics as the classical model if restricted to ``classical'' initial states without superpositions and diagonal observables. In the quantum model, however, one can also study general superpositions of initial states, and off-diagonal observables. Examples of these quantum circuits of permutations include the ``Rule 54'' cellular automaton~\cite{klobas2021exact,klobas2021exact2,klobas2021entanglement,gopalakrishnan2018operator,alba2019operator}, the CNOT gate~\cite{bertini2024exact,gopalakrishnan2018facilitated}, Yang-Baxter maps~\cite{gombor2022superintegrable}, and Goldilocks quantum cellular automata~\cite{hillberry2024integrability}. A common trait of all these examples is that whenever one does not consider classical states or observables, the ensuing dynamics looks qualitatively very similar to that of generic quantum systems. In fact, by means of numerical simulations Ref.~\cite{iaconis2021quantum} showed that a circuit that randomly selects between CNOT, SWAP, and $R_z$, produces wave functions that closely approximate the behaviour of fully Haar-random states. On the other hand, Ref.~\cite{lyons2023universal} demonstrated that a classical circuit with erasure errors exhibits qualitatively different physics when quantum gates are added. This again leads to the question of the role of quantumness in quantum many-body dynamics.

Here we aim to address this question more comprehensively by systematically comparing \emph{generic quantum} dynamics and \emph{generic classical} dynamics. To this end, we follow the fruitful approach proposed in Ref.~\cite{nahum2017quantum} and model generic quantum dynamics by considering a random unitary circuit (RUC): a quantum circuit where the gates are drawn at random from the set of unitary matrices. This produces a minimally structured quantum many-body system where only the constraints of unitarity and locality of the interactions are retained. Mirroring this approach in the classical realm, we then introduce \emph{random permutation circuits} (RPCs), where local updates are implemented by randomly selected elements of the set of permutations. This allows us to adapt the machinery of the Weingarten calculus~\cite{collins2022weingarten}, which has been very fruitful in RUCs (see e.g.\ Refs.~\cite{nahum2018operator,chan2018solution,vonKeyserlingk2018operator,zhou2019emergent,bertini2020scrambling}), to obtain exact results for averaged observables.

We consider the dynamics generated by RPCs both in the classical and in the quantum settings. The upshot of our analysis is that, in both cases, the qualitative features of RPCs are analogous to those of RUCs. More specifically, we show that the behaviour of decorrelator and mutual information in \emph{classical} RPCs matches that of OTOCs and entanglement in RUCs. In fact, also \emph{quantum} RPCs show the same phenomenology for OTOCs and entanglement whenever one considers off-diagonal observables, and initial states that are superpositions of computational-basis states. We also find exact quantitative correspondences between averaged quantities in RUCs and RPCs. For example, apart from an overall scale, we find that the averaged decorrelator in classical RPCs takes \emph{exactly} the same form as the averaged OTOCs in RUCs if one squares the local Hilbert space dimension. Similarly, we find that this is the case also for the averaged OTOCs between a diagonal and an off diagonal operator in quantum RPCs.

The rest of the paper is structured as follows. In Sec.~\ref{sec:setting} we introduce the framework of quantum circuits and explain our main results. In Sec.~\ref{sec:randomcircuits} we recall a few basic properties of RUCs and by analogy introduce RPCs. This is followed by Sec.~\ref{sec:QDwithP}, which contains a detailed discussion of the quantum dynamics generated by RPCs. Specifically, we present exact results on the averaged dynamical correlations, OTOCs, and purity. Then, in Sec.\ref{sec:class}, we focus on the classical dynamics of RPCs and present our conclusions in Sec.\ref{sec:conclusions}. Some further discussions, technical details, and proofs are reported in the Appendixes.

\section{Setting and Results}
\label{sec:setting}
We consider a one-dimensional chains of $2L$ qudits with local Hilbert space dimension $q$ evolving under a brickwork quantum circuit. The qudits are placed on a chain with lattice spacing $1/2$. Time is also discrete, and the evolution from time $t\in \mathbb N$ to $t+1$ is implemented by the unitary operator
\begin{equation}\label{eq:defTE}
  \mathbb{U}(t)=
  \smashoperator{\prod_{x\in\mathbb{Z}_L+1/2}} U(x,t+{1}/{2})_{x}
  \smashoperator{\prod_{x\in\mathbb{Z}_L}} U(x,t)_{x},
\end{equation}
so that the state of the system at time $t$, denoted by $\ket{\psi(t)}$, is given as
\begin{equation}
\label{eq:time_step}
\ket{\psi(t+1)} =  \mathbb{U}(t) \ket{\psi(t)} = \mathbb{U}(t) \cdots \mathbb{U}(0) \ket{\psi(0)}\,.
\end{equation}
Here the \emph{local gate} $U(x,t)\in{\rm End}(\mathbb C^{q^2})$ acts non-trivially on two qudits, and we introduced the notation $U(x,t)_{x}$ to denote the matrix in ${\rm End}(\mathbb C^{q^{2L}})$ acting as $U(x,t)$ on the qubits at positions $x$ and $x+1/2$ and as the identity operator elsewhere. Note that we do not assume any translational (in space or time) invariance and therefore, in general, $U(x,t) \neq U(x',t')$ for  $(x,t) \neq (x',t')$.  

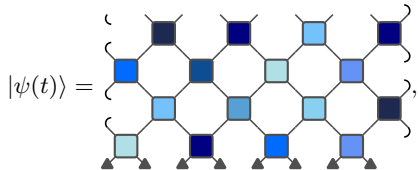
\begin{figure}
  \centering
  \begin{equation*}
    \ket{\psi(t)}=
    \begin{tikzpicture}[baseline={([yshift=-0.6ex]current bounding box.center)},scale=0.5]
      \prop{-1}{4}{myblue8}{}\prop{1}{4}{myblue2}{}\prop{3}{4}{myblue5}{}\prop{5}{4}{myblue2}{}
      \prop{-2}{3}{myblue3}{}\prop{0}{3}{myblue9}{}\prop{2}{3}{myblue1}{}\prop{4}{3}{myblue4}{}
      \prop{-1}{2}{myblue5}{}\prop{1}{2}{myblue6}{}\prop{3}{2}{myblue7}{}\prop{5}{2}{myblue8}{}
      \prop{-2}{1}{myblue1}{}\prop{0}{1}{myblue2}{}\prop{2}{1}{myblue3}{}\prop{4}{1}{myblue4}{}
      \foreach \x in {-3,...,4}{\inState{\x+0.5}{0.5}}
      \foreach \i in{1.5,3.5}{
        \draw[semithick,colLines] (5.5,\i+1) arc (-45:90:0.15);
        \draw[semithick,colLines] (5.5,\i) arc (45:-90:0.15);
        \draw[semithick,colLines] (-2.5,\i+1) arc (135:270:0.15);
        \draw[semithick,colLines] (-2.5,\i) arc (-135:-270:0.15);
      }
    \end{tikzpicture},
  \end{equation*}
  \caption{Time evolution of an initial product state, represented by dark triangles (cf.\ Appendix~\ref{app:diagrammatics} for a brief review of the diagrammatic notation).}
  \label{fig:statet}
\end{figure}

The evolution of the quantum circuit is conveniently represented diagrammatically, using the standard graphical representation of tensor networks, see, e.g.\ Ref.~\cite{cirac2020matrix}. In particular, depicting the local gate as
\begin{equation}\label{eq:localgate}
  U(x,t)=
  \begin{tikzpicture}[baseline={([yshift=-0.6ex]current bounding box.center)},scale=.75]
    \prop{0}{0}{colU}{}
  \end{tikzpicture},
\end{equation}
we can represent $\ket{\psi(t)}$ as in Fig.~\ref{fig:statet}, where we took $\ket{\psi(0)}$ to be a product state in space. In the figure, different shades represent different gates and the time direction runs from bottom to top, hence lower legs correspond to incoming indices (matrix row) and upper legs to outgoing indices (matrix column). Note that, following the standard conventions, when legs of different operators are joined together, a sum over the index of the corresponding local space is understood. For simplicity we assume periodic boundary conditions (unless otherwise specified). 

When studying the dynamics of quantum information one typically considers quantities involving $n\geq 2$ copies of $\ket{\psi(t)}$ and of its conjugate $\bra{\psi(t)}$. In this case the basic building block for the diagrammatic representation is
\be \label{eq:defReplicas}
    U^{(2n)}(x,t)\equiv (U(x,t) \otimes U^*(x,t))^{\otimes{n}} =
    \begin{tikzpicture}[baseline={([yshift=-0.6ex]current bounding box.center)},scale=.75]
        \prop{0}{0}{colU}{2n}
    \end{tikzpicture}\,,
\ee
where $(\cdot)^*$ denotes complex conjugation. In the tensor network language this representation is commonly referred to as the \emph{folded picture}~\cite{cirac2020matrix} (cf.\ Appendix~\ref{app:diagrammatics} for a brief review of the diagrammatic notation).

A special class of quantum circuits is that consisting of \emph{permutation unitaries}, which do not generate coherence in the local basis spanned by the qudit states $\{\ket s, s\in \mathbb Z_q\}$. Specifically, one requires $U(x,t)$ to be a permutation matrix in the basis $\{\ket{s_1, s_2}\equiv \ket{s_1}\otimes\ket{s_2}\}$. A circuit comprised of these gates always maps computational basis states to other computational basis states, i.e.\ it acts like a global permutation on the basis
\be
\mathcal B = \{\ket{s_1,\ldots,s_{2L}}, \qquad s_j=0,\ldots, q-1\}. 
\ee
The main motivation to study permutation unitaries is that they can be thought of as being derived from an underlying classical model.
Since in this case the time-evolution operator $\mathbb U(t)$ is a permutation, we can associate to this quantum model an equivalent classical reversible circuit composed of the same local permutation gates and evolving the classical configuration $(s_1,\dots,s_{2L})$ in precisely the same way as the quantum circuit evolves the basis state $\ket{s_1,\ldots,s_{2L}}$. While the time evolution of computational basis states is the same in both models, the key difference is that the quantum model permits superpositions. 
In contrast, in the classical model we can only consider probabilistic mixtures of basis states.
This can conveniently be represented by a density matrix (diagonal in the computational basis at all times)
\begin{equation}
	\rho(t+1) = \mathbb U(t) \rho(t)\mathbb U^\dagger(t).
	\label{eq:classical-time-evolution}
\end{equation}

To attain a universal characterisation of the dynamics we consider random circuits, where at each space-time point the local gates~\eqref{eq:localgate} are drawn at random from a specific ensemble. In particular we focus on two classes of random circuits (see also Sec.~\ref{sec:randomcircuits}). The first is the well-established family of random unitary circuits (RUCs)~\cite{nahum2017quantum, fisher2022random}, which has been extensively studied over the last few years, see e.g.\ Refs.~\cite{nahum2017quantum, vonKeyserlingk2018operator, nahum2018operator, chan2018solution, khemani2018operator, rakovszky2018diffusive, zhou2020entanglement, skinner2019measurement, wang2019barrier,skinner2019measurement,li2019measurement, chan_unitary-projective_2019, friedman2019spectral, bertini2018exact,chan2018spectral, garratt2021manybody, garratt2021local, jonay2018coarsegrained, zhou2019emergent}. We will use this class as a representative of generic, local quantum evolution.
The second class, which we dub \emph{random permutation circuits} (RPCs), is composed of the aforementioned permutation unitaries. In particular, we consider the random circuit where each $U(x,t)$ is drawn uniformly at random from the set of all permutations of $q^2$ elements. These circuits, and their corresponding classical models, can be seen as a caricature of the most general local classical reversible microscopic dynamics. Although quantum circuits of random permutations have been considered in the recent literature, see e.g.\ Refs.~\cite{iaconis2021quantum, feng2025dynamics}, to the best of our knowledge the class of Haar random permutation gates introduced here has not yet been studied.

A major breakthrough, which has led to deep insights into the dynamics of quantum information, has been the realisation that several probes of quantum information spreading, e.g.\ R\'enyi-2 entanglement entropy and OTOCs~\cite{vonKeyserlingk2018operator, nahum2018operator}, can be computed analytically when averaging over RUCs. Here we show that a similar averaging procedure leads to analytically tractable results also in the case of RPCs. Below we summarise our main results and discuss their implications. On the technical level, these results are obtained by finding a diagrammatic representation of the relevant quantities in terms of folded tensor networks, for which we derive closed systems of recurrence relations that can be solved in closed form. This approach was introduced in Ref.~\cite{bertini2020scrambling} and sidesteps the mapping to the classical spin model that is often used in random-circuit calculations, see e.g.\ Refs.~\cite{zhou2019emergent, nahum2018operator, fisher2022random}.  

\subsection{Results in the quantum setting}
To characterise the quantum dynamics generated by random permutations we study the behaviour of spatio-temporal correlations, the spreading of local operators, and the rate of entanglement production after a quantum quench, comparing them with results from random unitary circuits. The calculations are described in Sec.~\ref{sec:QDwithP}, while here we discuss the main results. 

First, in Sec.~\ref{sec:twopointCorr} we find that --- in contrast to the random-unitary case --- the correlations under random-permutation dynamics attain a non-zero value, and find a closed-form expression for the average. Specifically, we consider 
\begin{equation}
\label{eq:Cab}
  C_{\mu\nu}(x,t)=\frac{1}{\tr \1}
  \tr[\mathcal{O}_{\mu}(x,t)\mathcal{O}_{\nu}(0,0)],
\end{equation}
where $\mathcal{O}_{\mu}$ denotes a traceless, Hermitian operator acting on a single qudit, and is normalised as a generalised Pauli matrix, i.e.\ $\tr[\mathcal{O}_{\mu} \mathcal{O}_{\mu}^\dagger]=q$. We denote by $\mathcal{O}_{\mu}(x,0)$ the operator acting like $\mathcal{O}_{\mu}$ on the site $x$ and as the identity elsewhere, and by $\mathcal{O}_{\mu}(x,t)$ its Heisenberg picture evolution for $t$ steps.

\begin{figure}
   \centering
   \includegraphics[width=0.47\textwidth]{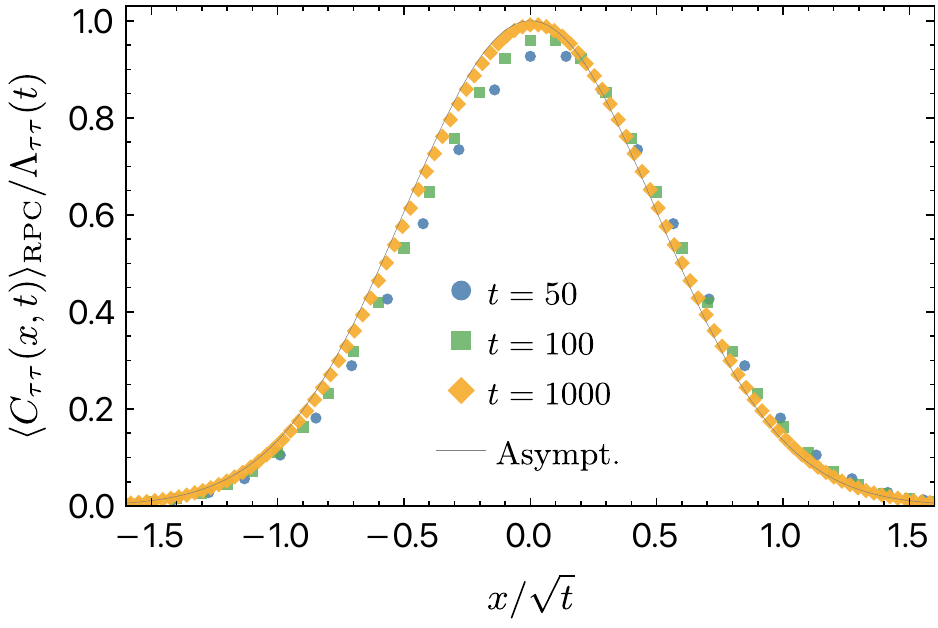}
   \caption{Rescaled correlation function  $\expval{C_{\tau\tau}(x,t)}_{\rpc}/\Lambda_{\tau\tau}(t)$ (cf.\ Eq.~\eqref{eq:avecorrresult}), and its asymptotic form (cf.\ Eq.~\eqref{eq:correlationasy}) versus $x/\sqrt{t}$. The prefactor $\Lambda_{\tau\tau}(t)$ is chosen to remove the overall correlation decay, $\Lambda_{\mu\nu}(t)=K_{\mu\nu}(4/q)^{2t}(1+1/q)^{-4t}/t^2$. The label $\tau$ refers to the clock operator, $\mathcal{O}_{\tau}\ket{s}=\ket{(s+1)\pmod q}$, and we chose $q=5$.
   }
   \label{fig:corr}
\end{figure}

The correlation function averaged over random permutations can be shown to fulfil a recurrence relation whose closed-form solution is reported in Eq.~\eqref{eq:avecorrresult}. As is shown in Appendix~\ref{sec:asyA}, for large space and time-scales the averaged correlation function takes a Gaussian form, rescaled by an exponentially decaying factor. In particular, denoting by $\expval{\cdot}_{\rpc}$ the average over random permutations, we have 
\be
\label{eq:correlationasy}
 \expval{C_{\mu\nu}(x,t)}_{\rpc} 
 \simeq \frac{K_{\mu \nu}}{t^2}
  \left(\frac{4/q}{(1+1/q)^2}\right)^{2t}
  e^{-2x^2/t},
\ee
where $K_{\mu\nu}$ is an observable-dependent constant that does not scale with $x$ or $t$, see Fig.~\ref{fig:corr}. Interestingly, the exact expression in Eq.~\eqref{eq:avecorrresult} is closely related to correlations in random unitary circuits. Indeed, although the averaged correlation function under random unitary dynamics is zero, we establish the following exact correspondence 
\begin{equation}\label{eq:pucorCorrs}
  \expval*{C^2_{\mu\nu}(x,t)}_{\ruc} \propto 
  \left.\expval*{C_{\mu\nu}(x,t)}_{\rpc}\right|_{q\mapsto q^2}\,,
\end{equation}
where $\expval{\cdot}_{\ruc}$ denotes the average over random unitaries, 
and the proportionality constant depends on the choice of $\mathcal{O}_{\mu}$ and $\mathcal{O}_{\nu}$ (see Sec.~\ref{sec:twopointCorr} for the details).

In Sec.~\ref{sec:OTOCs}, we provide a more precise characterisation of operator spreading by computing the OTOC between $\mathcal{O}_{\mu}$ and $\mathcal{O}_{\nu}$. Namely, we look at 
\begin{equation}
\label{eq:defOTOCs}
   {O}_{\mu\nu}(x,t)
   =\frac{1}{2}
  \frac{\tr[\smash{\left|[\mathcal{O}_{\mu}(x,t),\mathcal{O}_{\nu}(0,0)]\right|^2}]}{\tr\1},
\end{equation}
where $|\cdot|$ acting on a given operator $A$ should be understood as $|A|=\sqrt{AA^\dagger}$. Under permutation dynamics, the operators that are initially diagonal in the computational basis stay diagonal for all times, and therefore an OTOC between two such operators is identically zero for all $x$ and $t$. However, for generic $\mathcal{O}_{\mu}$, $\mathcal{O}_{\nu}$, Eq.~\eqref{eq:defOTOCs} is expected to asymptotically approach the generic step-like form. We show that the OTOC \emph{averaged} over random permutations can be exactly evaluated for a pair of operators, where only one is diagonal~\footnote{ Note that the averaged OTOC is of order one for small $x$, and the dynamics of computational basis states can be simulated efficiently. This implies that the OTOC profile can be efficiently estimated by sampling instances of the dynamics and averaging.}.  Surprisingly, the latter is found solving \emph{the same} system of recursive relations that also appear in the calculation of the averaged correlation function. This leads to the identity
\begin{equation} 
\label{eq:OTOCrandomperm}
 \!\!\!\expval{{O}_{\mathrm{d}\nu}(x,t)}_{\rpc}= (1-o_{\nu,2}) \left[\frac{\expval{C_{\mu\nu}(x,t)}_{\rpc}}{q o_{\mu} o_{\nu}}\right]\!\biggl |_{\alpha\mapsto1\!+\!q}\!\!, 
\end{equation}
where $\alpha=1+1/q$ is a parameter appearing in the solution of
$\expval{C_{\mu\nu}(x,t)}_{\rpc}$, and we replaced $\mu$ with $d$ to stress that
the first operator is diagonal. We also introduced the observable-dependent
constants 
\begin{equation} \label{eq:oalpha}
    o_\mu:=  \frac{1}{q^{{3}/{2}}}\sum_{i,j=0}^{q-1}\mel{i}{\mathcal{O}_{\mu}}{j}, 
\end{equation}
which parametrises the correlation functions, and 
\begin{equation} \label{eq:o2mu}
  o_{\mu, 2}:=\frac{1}{q}\sum_{i=0}^{q-1} \mel{i}{\mathcal{O}_{\mu}}{i}^2, 
\end{equation}
which measures the extent to which $\mathcal{O}_{\mu}$ is diagonal. The normalisation condition implies $0\le o_{{\mu},2}\le 1$, and $o_{\mu,2}=0$ whenever $\mathcal{O}_{\mu}$ is orthogonal to \emph{all} diagonal operators. For example, for a random Hermitian $\mathcal{O}_{\mu}$ we find $o_{\mu, 2}=1/q$. 

At large space-time scales Eq.~\eqref{eq:OTOCrandomperm} approaches a step-like form, signalling that operators spread with a butterfly velocity $v_B<1$~\cite{nahum2018operator, vonKeyserlingk2018operator}. In App.~\ref{sec:asyA}, we show that
\begin{equation}
\label{eq:OTOCasy}
  \frac{\expval{{O}_{\mathrm{d}\nu}(x,t)}_{\rpc}}{1- o_{\nu, 2}} \simeq  
  \Phi\left(\frac{v_B t+x }{\sigma(t)}\right)\Phi\left(\frac{v_B t-x }{\sigma(t)}\right),
\end{equation}
where we defined 
\begin{equation}\label{eq:defPhivbsigma}
\Phi(z)= \frac{1+\erf(z)}{2},\quad 
  v_B = \frac{q-1}{q+1},\quad 
  \sigma(t)= \frac{2\sqrt{q t}}{q+1}.
\end{equation}
Interestingly, by making the replacement $o_{\nu, 2}\mapsto 0$ and $q\mapsto q^2$, the r.h.s.\ of Eq.~\eqref{eq:OTOCasy} can be seen to coincide with the asymptotic form of the averaged OTOC in random unitary circuits~\cite{nahum2018operator}. In fact, in Sec.~\ref{sec:OTOCs} we establish the following \emph{exact} correspondence valid for any $x$ and $t$
\begin{equation}\label{eq:exactRUCsRPCs}
  \expval{{O}_{\mu\nu}(x,t)}_{\ruc} = \frac{1}{1-o_{\nu,2}}
  \left.\expval{{O}_{\mathrm{d}\nu}(x,t)}_{\rpc} \right|_{q\mapsto q^2} \,. 
\end{equation}

We remark that the prefactor in Eq.~\eqref{eq:exactRUCsRPCs} implies a quantitative difference with respect to the expected generic behaviour: for RUCs (or any sufficiently generic quantum circuit) the value inside the light cone is expected to be $1$. In the case of $\expval{{O}_{\mathrm{d}\nu}(x,t)}_{\rpc}$ the corresponding value is instead given by the weight of the off-diagonal part of $\mathcal{O_{\nu}}$, i.e.\ $1- o_{\nu, 2}\in[0,1]$ ($1-1/q$ for a random Hermitian operator). This can be intuitively explained by splitting $\mathcal{O}_{\nu}$ in its diagonal and off diagonal components, i.e.\
\begin{equation} \label{eq:opsplit}
  \mathcal{O}_{\nu}= \mathcal{O}_{{\rm d},\nu}+\mathcal{O}_{{\rm o},\nu}\,.
\end{equation}
Noting that this decomposition is preserved by the RPC evolution, and that diagonal operators commute, we then have  
\begin{equation}
  \mkern-8mu\expval{{O}_{{\rm d}\nu}(x,t)}_{\rpc}
   \!=\!\frac{1}{2}\mkern-2mu
  \expval*{\frac{\tr[\smash{\left|[\mathcal{O}_{d}(x,t),\mathcal{O}_{{\rm o}, \nu}(0,0)]\right|^2}]}{\tr\1}}_{\rpc}.\mkern-8mu
\end{equation}
The off-diagonal contribution to the OTOC (i.e.\ the quantity on the r.h.s.), behaves as an OTOC in random unitary circuits, but it is rescaled to account for the fact that the norm of $\mathcal{O}_{{\rm o},\nu}$ is strictly smaller than the norm of the full operator $\mathcal{O}_{\nu}$\footnote{In particular it is rescaled by $\tr[\mathcal{O}_{{\rm o}, \nu} \mathcal{O}_{{\rm o}, \nu}^\dagger]/q=1-o_{\nu, 2}$.}, which gives the overall prefactor $1-o_{\nu,2}$. The splitting in Eq.~\eqref{eq:opsplit} is conceptually similar to the one used in Ref.~\cite{khemani2018operator} to analyse random circuits with conservation laws: in that context one separates parallel and orthogonal components to the conserved densities. Rather than a conservation law, however, the fact that in our case the evolution preserves the split~\eqref{eq:opsplit} should be thought of as a dynamical constraint.   

Following the same logic, we then expect that the OTOC of two generic operators should still be asymptotically proportional to the r.h.s.\ of Eq.~\eqref{eq:OTOCasy} but the proportionality constant should change to 
\be
(1- o_{\nu, 2})\mapsto (1- o_{\mu, 2}o_{\nu, 2}),
\ee
which equals $1-1/q^2$ for random operators. This statement can be checked in the large $q$ limit, where one accordingly finds $\expval{{O}_{\mathrm{d}\nu}(x,t)}_{\rpc}=1-1/q^2+O(1/q^4)$ for $x \leq t$.

Finally, in Sec.~\ref{sec:entanglement} we consider entanglement growth.
While there exist special product states that remain unentangled (computational basis states and the uniform superposition over all basis states), generic non-fine-tuned low-entangled initial states exhibit linear growth of entanglement. To show this, we compute the average purity of half of the system after a quench from a product state $\ket{\psi(0)}$,
\begin{equation}
\begin{aligned}
  \expval{P(t)}_{\rpc}=
  \expval{\tr[\rho_{\rm half}(t)^2]}_{\rpc}, \\
 \rho_{\rm half}(t)= \tr_{\mathbb{Z}_L/2}\ketbra{\psi(t)}{\psi(t)}\,. 
\end{aligned}
\end{equation}
When the initial state is a random product state we find that at large $q$ the asymptotic behaviour of the purity ($1\ll t < L/2$) is given by  
\begin{equation}\label{eq:puritySummary}
  \expval{P(t)}_{\rpc} \simeq \left(\frac{4}{q}\right)^{2t}\,, \qquad q\gg4.
\end{equation}
This coincides with the result observed in random unitary circuits~\cite{nahum2018operator} with the replacement $q\mapsto q/2$ (cf.\ Table~\ref{tab:comparison}).

A common trait of all these results is that the qualitative behaviour of random-permutation circuits looks generic whenever one considers the dynamics of generic (i.e.\ not fine-tuned) operators or states. On the other hand, the existence of the ``special'' basis, in which the set of diagonal operators is mapped onto itself, represents a non-trivial dynamical constraint with measurable consequences, as evidenced by the prefactor in Eq.~\eqref{eq:exactRUCsRPCs}.

The correspondences between RUCs and RPCs are illustrated in \cref{tab:comparison}.  We note that there is no general correspondence, since correlations and OTOCs in RUC with $q$-dimensional qudits agree with those in RPCs with $q^2$-dimensional qudits, whereas the purities coincide when $q$-dimensional qudits in RUCs are paired with $2q$-dimensional qudits in RPCs. Moreover, these results only relate four-copy quantities on the RUC side with two- and four-copy quantities on the RPC side. Whether analogous exact correspondence can be found for larger numbers of copies remains an open question, but we expect the similarity to persist at least at the qualitative level. In the specific case of R\'enyi entropies
\be
S_\alpha(t)= \frac{1}{1-\alpha}\log\tr[\rho_{\rm half}(t)^\alpha], 
\ee
this can be argued using the inequality~\cite{wilming2019entanglement} 
\be
\frac{1}{2} S_2(t)\leq S_\alpha(t) \leq S_2(t),  \qquad \alpha\geq 2. 
\ee

\setlength{\tabcolsep}{8pt}
\begin{table}[t]
\caption{\label{tab:comparison}
    Relations between random quantum circuits (RUC) and random permutation circuits (RPC) (cf.\ Eqs.\ \eqref{eq:pucorCorrs}, \eqref{eq:exactRUCsRPCs}, and~\eqref{eq:puritySummary}).
    }
\begin{ruledtabular}
\begin{tabular}{lcccc}
	\textsc{Correlations}  & 
        $\expval*{C^2_{\mu\nu}(x,t)}_{\ruc} = \expval*{C_{\mu\nu}(x,t)}_{\rpc}\biggl |_{\substack{o_\mu \mapsto 1 \\ q\mapsto q^2}} $
        \\
	\textsc{OTOCs} & 
        $\expval{{O}_{\mu\nu}(x,t)}_{\ruc} = \expval{{O}_{\mathrm{d}\nu}(x,t)}_{\rpc} \biggl |_{\substack{o_{\nu , 2}\mapsto 0 \\ q\mapsto q^2}} \,$
        \\
	\textsc{Purity}  & 
    $\expval{P(t)}_{\ruc}\simeq \expval{P(t)}_{\rpc}|_{\substack{q\mapsto 2q}}$ \\ 
\end{tabular}
\end{ruledtabular}
\end{table}

\subsection{Results in the classical setting}
\label{sec:resultsclassical}

We can use the analytical results obtained from averaging over RPC also to characterise generic reversible classical dynamics. To this end, we first, in Sec.~\ref{sec:class1}, consider the \emph{decorrelator} $D(x,t)$ introduced in Ref.~\cite{Das2018,Liu2021a}, which is the classical analogue of the OTOC. It is defined by considering two copies of the system initialised in the same configuration. One then makes a local change at position $x=0$ in one of the two copies by applying a one-site traceless probability-conserving operator $\mathcal{O}_{\alpha}$ (e.g.\ a bit flip for $q=2$), and lets both copies undergo the same deterministic time evolution. The decorrelator measures the probability that the two configurations disagree at position $x$ and time $t$ upon taking a uniform average over all initial configurations (and random permutation gates). 

We find that, upon averaging over random permutations, the decorrelator becomes independent of the choice of $\mathcal{O}_{\alpha}$. More specifically, we obtain the exact relation 
\begin{equation}
  D(x,t)=\frac{1-{1}/{q}}{1-o_{\nu,2}}\expval{{O}_{d\nu}(x,t)}_{\rpc}.
\end{equation}
For large $x,t$, we can again use the asymptotic expansion~\eqref{eq:OTOCasy}, which yields
\be
    D(x,t) \simeq (1-1/q) \Phi\left(\!\frac{v_B t+x }{\sigma(t)}\!\right)\Phi\left(\!\frac{v_B t-x }{\sigma(t)}\!\right),
\ee
with $\Phi(z)$, $v_B$, and $\sigma(t)$ given in Eq.~\eqref{eq:defPhivbsigma}.

This immediately gives access to the \emph{damage spreading}, which quantifies how many sites will be flipped on average (i.e.\ the expected Hamming distance). This quantity is given by the sum of the decorrelator over all sites. Asymptotically it takes the form
\begin{equation} \label{eq:H}
  H(t) = \sum_x D(x,t) \rightarrow \left(1-\frac{1}{q}\right)2 v_B t.
\end{equation}
This expression has an intuitive interpretation: within the light cone the state is completely randomised, which means that the probability of the two copies agreeing on a particular site is $1/q$. The area in the light cone is $2v_Bt$, and thus Eq.~\eqref{eq:H} is just the expected number of disagreements in two randomly chosen computational basis states of length $2v_Bt$.

In Sec.~\ref{sec:class2} we study the spreading of information in the classical setting. We quantify the latter by computing the evolution of the mutual information $I(A:\bar A)$ between a subsystem $A$ and its complement $\bar A$~\cite{cover1999elements}. The mutual information is defined in terms of the entropies of the reduced systems 
\begin{equation}
  S_2(\rho)=-\log\tr(\rho^2),
\end{equation}
as
\begin{equation} 
\label{eq:mutualinfo}
  I(A:\bar A)=S_2(\rho_A)+S_2(\rho_{\bar A})-S_2(\rho),
\end{equation}
where $\rho_A=\tr_{\bar A}(\rho)$ and $\rho_{\bar A}=\tr_A(\rho)$. Since it is difficult to average $S_2$ over all circuits due to the presence of the $\log$, we use the customary annealed average approximation~\cite{nahum2018operator,vonKeyserlingk2018operator,zhou2019emergent}
\begin{equation}
  \expval{S_2(\rho)}_{\rpc}\approx-\log\expval{\tr(\rho^2)}_{\rpc},
\end{equation}
in terms of which we can define the annealed averaged mutual information
\begin{equation}\label{eq:mutualinfotilde}
  \tilde I(A:\bar A)=-\log\frac{\expval{\tr(\rho_A^2)}_{\rpc}\expval{\tr(\rho_{\bar A}^2)}_{\rpc}}{\expval{\tr(\rho^2)}_{\rpc}},
\end{equation}
which allows us to focus on the circuit-averaged purity.

If we initialise the system in the classical equivalent of a pure state, a single configuration, the mutual information remains zero at all times, because the entropy of the system (and any subsystem) is zero. Instead, we consider a mixed product initial state of the form 
\begin{equation} \label{eq:productinitialstatemixed}
  \rho(0)=\rho_0^{\otimes 2L},
\end{equation}
with $\rho_0$ diagonal, such that we can think of $\rho(0)$ as a classical mixture of computational basis states. We study the time evolution of this mixture under Eq.~\eqref{eq:classical-time-evolution}.  

Since the density matrix $\rho(t)$ (and, similarly, reduced states $\rho_{A,\bar{A}}(t)$) remains diagonal for all $t$, the calculation of the purity can be mapped to a double-replicated circuit (instead of a four-replica circuit for the usual quantum purity). This allows us to compute the short time ($t\leq L_A/2$) purity using a simple recurrence relation yielding 
\begin{equation}
    \tilde I(A:\bar{A})=-8t \log\left[\frac{q}{1+q}
    \left(\gamma+\frac{1}{q\, \gamma}\right)\right],
    \label{eq:MutualInfoResult}
\end{equation}
where we introduced the shorthand symbol $\gamma$ for the initial purity of a single-site density matrix
\begin{equation}
  \gamma=\tr(\rho_0^2),\qquad
  \frac{1}{q}\le \gamma\le 1.
\end{equation}
The mutual information is always zero for $\gamma =1/q$ and $\gamma=1$ corresponding to identity (totally mixed state) and a single computational basis states. For general states, however, we observe linear growth. 

\section{Random Circuits}
\label{sec:randomcircuits}

In this section we give a more detailed definition to the two classes of random circuits considered in this paper. We begin in Sec.~\ref{sec:ruc} by reviewing definition and basic properties of random unitary circuits. Then, in Sec.~\ref{sec:rpc}, we proceed by analogy to introduce random permutation circuits. 

\subsection{Random Unitary Circuits}
\label{sec:ruc}

In random unitary circuits the local gate $U(x,t)$ generating the evolution operator in Eq.~\eqref{eq:defTE} is sampled from the group of $q^2\times q^2$ unitary matrices, ${\rm U}(q^2)$, according to the Haar measure of the group (we denote it as ${\rm d}U$). This means that the average of a building block $U^{(n)}(x,t)$ reads 
\be
    \mathcal{U}^{(2n)} \equiv \expval*{U^{(2n)}}_{\ruc} = \smashoperator{\int\limits_{\mathrm U(q^2)}}
    (U \otimes U^*)^{\otimes{n}} {\rm d}U
    =
    \begin{tikzpicture}[baseline={([yshift=-0.6ex]current bounding box.center)},scale=.75]
        \prop{0}{0}{colIHaar}{2n}
    \end{tikzpicture}.
    \label{eq:explicitaverageruc}
\ee
Because of the invariance of the Haar measure, this object is a \emph{projector}, i.e.\ $(\mathcal{U}^{(2n)})^2=\mathcal{U}^{(2n)} $. In particular, it projects on the space spanned by the states
\be
\{\ket*{\sigma_{j}}\otimes \ket*{\sigma_{j}}\}_{j=1}^{n!} \equiv \{\ket{\sigma_{j} \sigma_{j}}\}_{j=1}^{n!},
\label{eq:doubleperm}
\ee
where $\{\ket*{\sigma_{j}}\}$ is in one-to-one correspondence with ${\rm S}(n)$, the set of permutations of $n$ objects. Namely, for each permutation $\sigma_{j}\in{\rm S}(n)$ one defines a state $\ket*{\sigma_{j}}$ with the following coefficients in the computational basis of $2n$ qudits ($\ket{s_1r_2\ldots s_nr_n}_{s_j,r_j=1}^q$)
\be
\braket{s_1 r_1\ldots s_{n} r_n}{\sigma_{j}} = \prod_{m=1}^{n}
    \delta_{s_{m}, r_{\sigma(m)}}\,.
\label{eq:coefficientsperm}
\ee 
Note that these states are \emph{not} orthogonal, and their scalar product is given by~\cite{zhou2019emergent} 
\be
\braket{\sigma_{i}}{\sigma_{j}} = q^{\mathfrak c(\sigma_{i}\sigma^{-1}_{j})},
\label{eq:overlapperm}
\ee
where $\mathfrak c(\sigma)$ denotes the number of cycles in the permutation $\sigma$. This means that also the states \eqref{eq:doubleperm} are not orthogonal and we denote their Gram matrix by 
\be
[G_{\rm U}^{(2n)}]_{ij}=\braket{\sigma_{i}\sigma_{i}}{\sigma_{j}\sigma_{j}} =\braket{\sigma_{i}}{\sigma_{j}}^2.
\label{eq:Gramperm}
\ee
Whenever the vectors \eqref{eq:doubleperm} are linearly independent (always true for fixed $n$ and large enough $q$) this matrix is positive definite. Therefore, proceeding by standard orthogonalisation, we can express the projector $\mathcal{U}^{(2n)}$ as 
\begin{equation}
	\mathcal{U}^{(2n)}=\sum_{i,j=1}^{n!}  [(G_{\rm U}^{(2n)})^{-1}]_{ij}  \ketbra{\sigma_{i}\sigma_{i}}{\sigma_{j}\sigma_{j}}. 
	\label{eq:U2n}
\end{equation}
The inverse of the Gram matrix appearing in this expression is known in the literature as the \emph{Weingarten matrix}, or \emph{Weingarten function}~\cite{collins2022weingarten}, therefore we denote it by $W_{\rm U}^{(2n)}$. Note that in the case at hand $[W_{\rm U}^{(2n)}]_{ij}$ only depends on $\sigma_{i}\sigma_{j}^{-1}$. For example, for $n=2$ we have two permutations of two elements, the identity permutation $\sigma_1$, and the transposition $\sigma_2$,
\begin{equation}
  \sigma_1(1)=1,\,\sigma_1(2)=2,\qquad
  \sigma_2(1)=2,\,\sigma_2(2)=1.
\end{equation}
which correspond to the invariant states $\ket{\sigma_{1}}$ and $\ket{\sigma_2}$ (see App.~\ref{app:diagrammatics} for their explicit form). Their overlaps imply the following Weingarten matrix
\be 
W_{\rm U}^{(4)}= \frac{1}{q^2(q^4-1)}
\begin{bmatrix}
\phantom{-}q^2  & -1\\
-1 & \phantom{-}q^2\\ 
\end{bmatrix}\,,
\ee
which means that applying $\mathcal{U}^{(4)}$ to tensor products of the states 
\begin{equation}
  \label{eq:squarecircdef}
  \ket*{\circleSA^{(4)}}:=\frac{1}{q}\ket*{\sigma_{1}}= \begin{tikzpicture}[baseline={([yshift=-0.6ex]current bounding box.center)},scale=0.75]
    \gridLine{-0.3}{-0.3}{-0.3}{0.45}
    \circle{-0.3}{-0.3}
    \node[scale=0.7] at (-0.05,-.35) {$4$};
  \end{tikzpicture}, \quad
  \ket*{\squareSA^{(4)}}:=\frac{1}{q}\ket{\sigma_{2}} =   \begin{tikzpicture}[baseline={([yshift=-0.6ex]current bounding box.center)},scale=0.75]
    \gridLine{-0.3}{-0.3}{-0.3}{0.45}
    \square{-0.3}{-0.3}
    \node[scale =0.7] at (-0.05,-.35) {$4$};
  \end{tikzpicture},
\end{equation}
gives 
\begin{equation} \label{eq:aveUonstates}
  \begin{gathered}
    \mathcal{U}^{(4)}\ket*{\circleSA\,\circleSA}=\ket*{\circleSA\,\circleSA}, \qquad \mathcal{U}^{(4)}\ket*{\squareSA\,\squareSA}=\ket*{\squareSA\,\squareSA},\\
    \mathcal{U}^{(4)}\ket*{\circleSA\,\squareSA}=\mathcal{U}^{(4)}\ket*{\squareSA\,\circleSA}= \frac{{q}}{1+q^2} \left(\ket*{\circleSA\,\circleSA}+\ket*{\squareSA\,\squareSA}\right),
  \end{gathered}
\end{equation}
where the relations on the first line are a direct consequence of the unitarity of the dynamics and hold for any unitary gate or average of unitary gates, while the second depends on the specific form of the averaged gate~\eqref{eq:U2n}. Note that in Eq.~\eqref{eq:squarecircdef} we explicitly reported in a superscript the number of replicas in which the states on the l.h.s.\ are defined. Instead, in Eq.~\eqref{eq:aveUonstates} we suppressed these superscripts as the number of replicas is already reported in the averaged gate. In the rest of the paper we will follow this convention and suppress the superscript on states every time that the number of replicas is clear from the context. 

Using the diagrammatic representation in Eqs.~\eqref{eq:explicitaverageruc} and \eqref{eq:squarecircdef} we can depict these relations as 
\begin{equation}\label{eq:UdEval2}
  \begin{gathered}
\begin{tikzpicture}[baseline={([yshift=-0.6ex]current bounding box.center)},scale=0.75]
    \prop{0}{0}{colIHaar}{4}
    \circle{-0.5}{-0.5}
    \circle{0.5}{-0.5}
  \end{tikzpicture}=
  \begin{tikzpicture}[baseline={([yshift=-0.6ex]current bounding box.center)},scale=0.75]
    \gridLine{-0.3}{-0.3}{-0.3}{0.45}
    \gridLine{0.3}{-0.3}{0.3}{0.45}
    \circle{-0.3}{-0.3}
    \circle{0.3}{-0.3}
  \end{tikzpicture},\qquad
  \begin{tikzpicture}[baseline={([yshift=-0.6ex]current bounding box.center)},scale=0.75]
    \prop{0}{0}{colIHaar}{4}
    \square{-0.5}{-0.5}
    \square{0.5}{-0.5}
  \end{tikzpicture}=
  \begin{tikzpicture}[baseline={([yshift=-0.6ex]current bounding box.center)},scale=0.75]
    \gridLine{-0.3}{-0.3}{-0.3}{0.45}
    \gridLine{0.3}{-0.3}{0.3}{0.45}
    \square{-0.3}{-0.3}
    \square{0.3}{-0.3}
  \end{tikzpicture},\\
\begin{tikzpicture}[baseline={([yshift=-0.6ex]current bounding box.center)},scale=0.75]
    \prop{0}{0}{colIHaar}{4}
    \square{-0.5}{-0.5}
    \circle{0.5}{-0.5}
  \end{tikzpicture}=
  \begin{tikzpicture}[baseline={([yshift=-0.6ex]current bounding box.center)},scale=0.75]
    \prop{0}{0}{colIHaar}{4}
    \circle{-0.5}{-0.5}
    \square{0.5}{-0.5}
  \end{tikzpicture}=
 \frac{{q}}{1+q^2} 
  \left(
    \begin{tikzpicture}[baseline={([yshift=-0.6ex]current bounding box.center)},scale=0.75]
    \gridLine{-0.3}{-0.3}{-0.3}{0.45}
    \gridLine{0.3}{-0.3}{0.3}{0.45}
    \circle{-0.3}{-0.3}
    \circle{0.3}{-0.3}
  \end{tikzpicture}+
  \begin{tikzpicture}[baseline={([yshift=-0.6ex]current bounding box.center)},scale=0.75]
    \gridLine{-0.3}{-0.3}{-0.3}{0.45}
    \gridLine{0.3}{-0.3}{0.3}{0.45}
    \square{-0.3}{-0.3}
    \square{0.3}{-0.3}
  \end{tikzpicture}\right).
  \end{gathered}
\end{equation}
Since $\mathcal{U}^{(4)}$ is symmetric these relations also hold when the diagrams are flipped top to bottom.

\subsection{Random Permutation Circuits}
\label{sec:rpc}
When considering circuits of random permutations we sample the local gate $U(x,t)$ in Eq.~\eqref{eq:defTE} from a unitary representation of the permutation group. These matrices implement permutations in a special basis, which we take to be the computational basis of two qudits 
\begin{equation} \label{eq:computationalbasis}
  \{\ket{s}\otimes \ket{r} \equiv \ket{s,r}\}_{s,r=1}^q\,.
\end{equation}
Recalling our discussion in Sec.~\ref{sec:setting}, this is the special basis in which the circuit acts classically. More formally, we define the set 
\begin{equation} \label{eq:permutations}
  {\rm S}(q^2) = \{U\in {\rm U}(q^2): U \ket{s,r}= \ket{ f(s,r)}\},
\end{equation}
where $f$ is an invertible endomorphism of $\mathbb{Z}_{q}\times \mathbb{Z}_{q}$. It is easy to see that, when equipped with the standard matrix product, ${\rm S}(q^2)$ is indeed a finite group of order $q^2!$. 

Since ${\rm S}(q^2)$ is a finite group its Haar measure is the flat one (also referred to as \emph{counting measure}), see, e.g.\ Ref.~\cite{collins2022weingarten}. Therefore, the averaged tensor-product gate is written as 
\begin{equation}
  \label{eq:permave}
  \mathcal{Q}^{(n)}\equiv \expval*{U^{(n)}}_{\rpc} =\frac{1}{q^2!}
  \smashoperator[r]{\sum_{U\in {\rm S}(q^2)}}U^{\otimes{n}}=
  \begin{tikzpicture}[baseline={([yshift=-0.6ex]current bounding box.center)},scale=.75]
    \prop{0}{0}{colPerm}{n}
  \end{tikzpicture},
\end{equation}
where we have assumed that the gates $U$ are written in the computational basis \eqref{eq:computationalbasis} and thus are real. Note that the averaged gate in Eq.~\eqref{eq:permave} is defined for any positive integer in the exponent $n$, not only for even numbers as in Eq.~\eqref{eq:explicitaverageruc}. 

Similarly to the random unitary case the averaged gate in Eq.~\eqref{eq:permave} is a projector into the space spanned by appropriate invariant vectors. These vectors are 
\begin{equation} \label{eq:doublepart}
  \{\ket*{\pi_{j}}\otimes \ket*{\pi_{j}}\}_{j=1}^{B_n} 
  \equiv \{\ket*{\pi_{j}\pi_{j}}\}_{j=1}^{B_n},
\end{equation}
where $\ket*{\pi_{j}}$ 
are in a one-to-one correspondence with the \emph{partitions} of $\mathbb{Z}_{n}$, where $j$ is (\emph{a priori} non-uniquely) specifying the partition. We recall that a partition $\pi_{j}$ of $\mathbb{Z}_n$ is defined as
\be
\pi_{j}=\{\!\{i_{1,1},\ldots,i_{1,\ell_1}\},\ldots
\{i_{k,1},\ldots,i_{k,\ell_k}\}\!\}
\label{eq:partition}
\ee
with 
\begin{align}
\sum_{m=1}^{k} \ell_m=n,\qquad 
\bigcup_{m=1}^{k} \{i_{m,1},\ldots,i_{m,\ell_m}\}=\mathbb{Z}_n.
\end{align}
The number $B_n$ of partitions of $\mathbb{Z}_n$ is known as the \emph{Bell number}~\cite{bell1934exponential,rota1964number}, and fulfils the following recurrence relation
\begin{equation}
  B_0=B_1=1,\qquad
  B_n=\sum_{i=0}^{n-1}\binom{n-1}{i} B_i.
\end{equation}
The vector ${\ket{\pi_{j}}\in \mathbb{C}^{q^n}}$ corresponding to the partition $\pi_{j}$ is defined by the following coefficients in the computational basis 
\be
\braket{s_1 s_2\ldots s_n}{\pi_{j}} = \prod_{m=1}^{k}
  \left(
    \prod_{r=2}^{\ell_m}
    \delta_{s_{i_{m, r-1}},s_{i_{m, r}}}
\right)\,.
\label{eq:coefficientspart}
\ee 
We see that set of states $\{\ket*{\pi_{2n,j}}\}$ includes $\{\ket*{\sigma_{2n,j}}\}$, the one considered in the previous subsection, as a subset. Indeed, Eq.~\eqref{eq:coefficientsperm} can be obtained as a special case of Eq.~\eqref{eq:coefficientspart}~\footnote{One has to set $k=n$, $\ell_1,\ldots,\ell_n=2$ and $i_{m, 1}=m$ and $i_{m,2}=\sigma(m)$}.
One can explicitly enumerate all partitions and the corresponding vectors for the first few $n$. In particular, in Appendix~\ref{app:partitions} we report them for $n=1,2,3,4$, which correspond respectively to sets of $1,2,5$, and $15$ vectors. 

Note that the appearance of partition states in Eq.~\eqref{eq:doublepart} can be understood by recalling that $U$ acts as a one-to-one map between two-site basis states $\ket{s_1 s_2}$,
\begin{equation}
  U\ket{s_1 s_2} = \ket{f(s_1 s_2)},\qquad
  f,f^{-1}\in \mathrm{End}(\mathbb{Z}_{q}\otimes\mathbb{Z}_{q}),
\end{equation}
and therefore also $U^{\otimes k}$ with an arbitrary $k\in\mathbb{N}$ bijectively maps the set of replicated basis states $\{\ket{s_1 s_2}^{\otimes k} \}_{0\le s_1,s_2\le q-1}$ to itself. From here it directly follows that all the states of the form~\eqref{eq:doublepart} are invariant under $U^{\otimes n}$,
\begin{equation}
  U^{\otimes n}\ket{\pi_j\pi_j}=\ket{\pi_j\pi_j}.
\end{equation}
We remark that this holds for \emph{any} permutation unitary $U$, and therefore is not a consequence of averaging.

As before, to arrive to an explicit expression of $\mathcal{Q}^{(n)}$ in terms of $\{\ket*{\pi_{j} \pi_{j}}\}$ one proceeds to orthogonalise the set. This leads to the following expression 
\be
\mathcal{Q}^{(n)} = \sum_{i,j=1}^{B_n}  [W_{\rm P}^{(n)}]_{ij}  \ketbra{\pi_{i} \pi_{i}}{\pi_{j} \pi_{j}},
\ee
where the Weingarten function of the permutation group is the inverse of the Gram matrix of the double partition states
\be
[G_{\rm P}^{(n)}]_{ij}= \braket{\pi_{i} \pi_{i}}{\pi_{j} \pi_{j}} = \braket{\pi_{i}}{\pi_{j}}^2\,.
\ee
For example, for $n=2$ the Weingarten function is 
\be
\label{eq:Q2Eval}
W_{\rm P}^{(2)}= \frac{1}{q^2(q^2-1)} \begin{bmatrix}
\phantom{-}q^2  & -1\\
-1 & \phantom{-}1\\ 
\end{bmatrix}\,.
\ee
This means that, applying $\mathcal{Q}^{(2)}$ to the tensor products of the states 
\begin{equation}
 \begin{aligned}
 \label{eq:flatcircdef}
   \ket*{\circleSA^{(2)}}&:=\frac{1}{\sqrt{q}}\ket*{\sigma_{1}}=\frac{1}{\sqrt{q}}\ket{\pi_{1}} =  \begin{tikzpicture}[baseline={([yshift=-0.6ex]current bounding box.center)},scale=0.75]
    \gridLine{-0.3}{-0.3}{-0.3}{0.45}
    \circle{-0.3}{-0.3}
     \node[scale =0.7] at (-0.05,-.35) {$2$};
  \end{tikzpicture},\\
   \ket*{\flatSA^{(2)}}&:=\frac{1}{q}\ket{\pi_{2}}=\begin{tikzpicture}[baseline={([yshift=-0.6ex]current bounding box.center)},scale=0.75]
    \gridLine{-0.3}{-0.3}{-0.3}{0.45}
    \flat{-0.3}{-0.3}
    \node[scale =0.7] at (-0.05,-.35) {$2$};
  \end{tikzpicture},
\end{aligned}
\end{equation}
we find 
\begin{equation}
  \begin{gathered}
    \mathcal{Q}^{(2)}\ket*{\circleSA\,\circleSA}=\ket*{\circleSA\,\circleSA}, \qquad \mathcal{Q}^{(2)}\ket*{\flatSA\,\flatSA}=\ket*{\flatSA\,\flatSA},\\
    \mathcal{Q}^{(2)}\ket*{\circleSA\,\flatSA}=\mathcal{Q}^{(2)}\ket*{\flatSA\,\circleSA}= \frac{\sqrt{q}}{1+q} \left(\ket*{\circleSA\,\circleSA}\!+\!\!\!\ket*{\flatSA\,\flatSA}\right),
  \end{gathered}
\end{equation}
where, as before, the first relation on the first line is a direct consequence of the unitarity of the dynamics, while the second reflects the deterministic nature of permutation circuits. Therefore, they hold for any permutation circuit and any average of them. Instead, the relation on the third line depends on the specific form of $\mathcal{Q}^{(2)}$.

Using the diagrammatic representation introduced above these relations can be depicted as 
\begin{equation}\label{eq:WdEval2}
  \begin{gathered}
\begin{tikzpicture}[baseline={([yshift=-0.6ex]current bounding box.center)},scale=0.75]
    \prop{0}{0}{colPerm}{2}
    \circle{-0.5}{-0.5}
    \circle{0.5}{-0.5}
  \end{tikzpicture}=
  \begin{tikzpicture}[baseline={([yshift=-0.6ex]current bounding box.center)},scale=0.75]
    \gridLine{-0.3}{-0.3}{-0.3}{0.45}
    \gridLine{0.3}{-0.3}{0.3}{0.45}
    \circle{-0.3}{-0.3}
    \circle{0.3}{-0.3}
  \end{tikzpicture},\qquad
  \begin{tikzpicture}[baseline={([yshift=-0.6ex]current bounding box.center)},scale=0.75]
    \prop{0}{0}{colPerm}{2}
    \flatLD{-0.5}{-0.5}
    \flatRD{0.5}{-0.5}
  \end{tikzpicture}=
  \begin{tikzpicture}[baseline={([yshift=-0.6ex]current bounding box.center)},scale=0.75]
    \gridLine{-0.3}{-0.3}{-0.3}{0.45}
    \gridLine{0.3}{-0.3}{0.3}{0.45}
    \flat{-0.3}{-0.3}
    \flat{0.3}{-0.3}
  \end{tikzpicture},\\
\begin{tikzpicture}[baseline={([yshift=-0.6ex]current bounding box.center)},scale=0.75]
    \prop{0}{0}{colPerm}{2}
    \flatLD{-0.5}{-0.5}
    \circle{0.5}{-0.5}
  \end{tikzpicture}=
  \begin{tikzpicture}[baseline={([yshift=-0.6ex]current bounding box.center)},scale=0.75]
    \prop{0}{0}{colPerm}{2}
    \circle{-0.5}{-0.5}
    \flatRD{0.5}{-0.5}
  \end{tikzpicture}=
  \frac{\sqrt{q}}{1+q}
  \left(
  \begin{tikzpicture}[baseline={([yshift=-0.6ex]current bounding box.center)},scale=0.75]
    \gridLine{-0.3}{-0.3}{-0.3}{0.45}
    \gridLine{0.3}{-0.3}{0.3}{0.45}
    \circle{-0.3}{-0.3}
    \circle{0.3}{-0.3}
  \end{tikzpicture}+
  \begin{tikzpicture}[baseline={([yshift=-0.6ex]current bounding box.center)},scale=0.75]
    \gridLine{-0.3}{-0.3}{-0.3}{0.45}
    \gridLine{0.3}{-0.3}{0.3}{0.45}
    \flat{-0.3}{-0.3}
    \flat{0.3}{-0.3}
  \end{tikzpicture}
\right).
  \end{gathered}
\end{equation}
Once again, the symmetric structure of $\mathcal{Q}^{(2)}$ implies that these relations also hold when the diagrams are flipped top to bottom. 

Interestingly, these equations are exactly equivalent to Eq.~\eqref{eq:UdEval2} upon performing the mapping $q\mapsto q^2$. As we see in the rest of the paper, this leads to formal correspondences between quantities computed in random unitary and random permutation circuits.

\section{Quantum Dynamics}
\label{sec:QDwithP}

In this section we characterise the quantum dynamics generated by random
permutation gates and compare it with the dynamics of random unitary circuits.
Specifically in Sec.~\ref{sec:twopointCorr} we study spatio-temporal
correlations. In Sec.~\ref{sec:OTOCs} we characterise operator spreading by
computing OTOCs, and in Sec.~\ref{sec:entanglement} we consider entanglement
dynamics.

{As we discussed in Sec.~\ref{sec:setting} the results obtained here show that the dynamics generated by circuits of random permutations for generic operators and states have the same qualitative features as those of real quantum circuits.}

\subsection{Two-point correlation functions}
\label{sec:twopointCorr}
Let us begin considering the two-point correlation functions over the infinite temperature state, $C_{\mu\nu}(x,t)$, defined in Eq.~\eqref{eq:Cab}. Even though $x$ takes half-integer values, here we for clarity choose it to be an integer --- the case of $x$ half-odd-integer is completely analogous.

The average of $C_{\mu\nu}(x,t)$ over random permutations can be written in terms of $\mathcal{Q}^{(2)}$ as (cf.\ Appendix~\ref{app:diagrammatics})
\begin{equation}
 \label{eq:averaged_Cab}
  \mkern-4mu
  \expval{C_{\mu\nu}(x,t)}_{\rpc}=\mkern-25mu
  \begin{tikzpicture}[baseline={([yshift=-2.2ex]current bounding box.center)},scale=0.5]
    \prop{0}{0}{colPerm}{2}
    \foreach \x in {-1,1}{\prop{\x}{1}{colPerm}{2}}
    \foreach \x in {-2,0,2}{\prop{\x}{2}{colPerm}{2}}
    \foreach \x in {-3,-1,...,3}{\prop{\x}{3}{colPerm}{2}}
    \foreach \x in {-4,-2,0,2}{\prop{\x}{4}{colPerm}{2}}
    \foreach \x in {-5,-3,-1,1}{\prop{\x}{5}{colPerm}{2}}
    \foreach \x in {-6,-4,-2,0}{\prop{\x}{6}{colPerm}{2}}
    \foreach \x in {-5,-3,-1}{\prop{\x}{7}{colPerm}{2}}
    \foreach \x in {-4,-2}{\prop{\x}{8}{colPerm}{2}}
    \foreach \x in {-3}{\prop{\x}{9}{colPerm}{2}}
    \foreach \x in {0,...,6}{\circle{-0.5-\x}{-0.5+\x}}
    \foreach \x in {7,...,9}{\circle{-13.5+\x}{-0.5+\x}}
    \foreach \x in {1,...,3}{\circle{0.5+\x}{-0.5+\x}}
    \foreach \x in {4,...,10}{\circle{7.5-\x}{-0.5+\x}}
    \blackcircle{0.5}{-0.5}
    \blackcircle{-3.5}{9.5}
    \node at (1,-1) {\scalebox{0.9}{$\mathcal{O}_{\mu}$}};
    \node at (-3.1,9.9) {\scalebox{0.9}{$\mathcal{O}_{\nu}$}};
    \draw[black,|<->|] (-3.5,-1.5) -- (0.5,-1.5) node[midway,above] {$2x$};
    \draw[black,|<->|] (4,-0.5) -- (4,9.5) node[pos=0.75,left] {$2t$};
  \end{tikzpicture},
  \mkern-4mu
\end{equation}
where in the diagram above we set $t=5$, $x=2$ and introduced the operator-dependent
states,
\begin{equation}
  \ket*{\blackcircleSA{\mu}^{(2)}} = \mathcal{O}_{\mu}\otimes\1 \ket*{\circleSA^{(2)}} 
  = \begin{tikzpicture}[baseline={([yshift=-0.6ex]current bounding box.center)},scale=0.75]
    \gridLine{0.5}{-0.5}{0.5}{0.25}
    \blackcircle{0.5}{-0.5}
    \node at (1,-0.5) {\scalebox{0.9}{$\mathcal{O}_{\mu}$}};
  \end{tikzpicture}  \,.
  \label{eq:blackcirclestate}
\end{equation} 
Since $\mathcal{Q}^{(2)}$ is a projector on the subspace spanned by products of $\ket{\circleSA^{(2)}}$ and $\ket{\flatSA^{(2)}}$ (cf.~\eqref{eq:flatcircdef}), the states corresponding to the operators $\mathcal{O}_{\alpha}$ and $\mathcal{O}_{\beta}$ are projected onto the space spanned by $\ket{\circleSA}$ and $\ket{\flatSA}$. In particular, by computing the overlaps (cf.\ Eq.~\eqref{eq:oalpha})  
\begin{equation}
\begin{aligned}
  \braket*{\flatSA^{(2)}}{\blackcircleSA{\mu}^{(2)}}&=q^{-\frac{3}{2}}
  \sum_{i,j=0}^{q-1}\mel{i}{\mathcal{O}_{\mu}}{j} = {o_\mu}
  ,  \\
  \braket*{\circleSA^{(2)}}{\blackcircleSA{\alpha}^{(2)}}&=
  q^{-1}\tr[\mathcal{O}_\alpha]=0,
     \end{aligned}
\end{equation}
and using the first line of Eq.~\eqref{eq:WdEval2} we find 
\begin{equation}\label{eq:CgenCflat}
  \mkern-8mu
 \expval{C_{\mu\nu}(x,t)}_{\rpc} =
  \frac{o_{\mu} o_{\nu} }{\left(1-\frac{1}{q}\right)^2}
  \left(\expval{C_{\SflatSA\SflatSA}(x,t)}_{\rpc}  -\frac{1}{q}\right),
  \mkern-8mu
\end{equation}
where by $o_{\mu}$ we denote the rescaled sum of matrix elements of $\mathcal{O}_{\mu}$ (cf.\ Eq.~\eqref{eq:oalpha}). Eq.~\eqref{eq:CgenCflat} shows that all non-trivial information is contained in $\expval{C_{\SflatSA\SflatSA}(x,t)}_{\rpc}$. 

Crucially, we now show that the latter can be computed exactly by writing and solving a system of recurrence relations. First we introduce the following two families of auxiliary functions
\begin{equation}
\label{eq:wk}
  w_{k}(x,y) :=
  \begin{tikzpicture}[baseline={([yshift=-0.6ex]current bounding box.center)},scale=0.5]
    \prop{0}{0}{colPerm}{2}
    \foreach \x in {-1,1}{\prop{\x}{1}{colPerm}{2}}
    \foreach \x in {-2,0,2}{\prop{\x}{2}{colPerm}{2}}
    \foreach \x in {-3,-1,...,3}{\prop{\x}{3}{colPerm}{2}}
    \foreach \x in {-4,-2,0,2}{\prop{\x}{4}{colPerm}{2}}
    \foreach \x in {-5,-3,-1,1}{\prop{\x}{5}{colPerm}{2}}
    \foreach \x in {-6,-4,-2,0}{\prop{\x}{6}{colPerm}{2}}
    \foreach \x in {-5,-3,-1}{\prop{\x}{7}{colPerm}{2}}
    \foreach \x in {-4,-2}{\prop{\x}{8}{colPerm}{2}}
    \foreach \x in {-3}{\prop{\x}{9}{colPerm}{2}}
    \foreach \x in {0,...,6}{\circle{-0.5-\x}{-0.5+\x}}
    \foreach \x in {7,...,9}{\circle{-13.5+\x}{-0.5+\x}}
    \foreach \x in {2,...,3}{\circle{0.5+\x}{-0.5+\x}}
    \foreach \x in {4,...,10}{\circle{7.5-\x}{-0.5+\x}}
    \flatRD{0.5}{-0.5}
    \flatRD{1.5}{0.5}
    \flatRD{-3.5}{9.5}
    \draw[semithick,decorate,decoration={brace}] (-6.65,6.65) -- (-3.65,9.65) node[midway,xshift=-5pt,yshift=5pt] {$x$};
    \draw[semithick,decorate,decoration={brace}] (-2.35,9.65) -- (3.65,3.65) node[midway,xshift=5pt,yshift=5pt] {$y$};
    \draw[semithick,decorate,decoration={brace}] (1.65,0.35) -- (0.65,-0.65) node[midway,xshift=5pt,yshift=-5pt] {$k$};
    \draw[semithick,decorate,decoration={brace}] (3.65,2.35) -- (2.65,1.35) node[midway,xshift=9pt,yshift=-5pt] {$x\!-\!k$};
  \end{tikzpicture},
\end{equation}
and 
\begin{equation}
  v_{k}(x,y) :=
  \begin{tikzpicture}[baseline={([yshift=-0.6ex]current bounding box.center)},scale=0.5]
    \foreach \x in {-1}{\prop{\x}{1}{colPerm}{2}}
    \foreach \x in {-2,0,2}{\prop{\x}{2}{colPerm}{2}}
    \foreach \x in {-3,-1,...,3}{\prop{\x}{3}{colPerm}{2}}
    \foreach \x in {-4,-2,0,2}{\prop{\x}{4}{colPerm}{2}}
    \foreach \x in {-5,-3,-1,1}{\prop{\x}{5}{colPerm}{2}}
    \foreach \x in {-6,-4,-2,0}{\prop{\x}{6}{colPerm}{2}}
    \foreach \x in {-5,-3,-1}{\prop{\x}{7}{colPerm}{2}}
    \foreach \x in {-4,-2}{\prop{\x}{8}{colPerm}{2}}
    \foreach \x in {-3}{\prop{\x}{9}{colPerm}{2}}
    \foreach \x in {1,...,6}{\circle{-0.5-\x}{-0.5+\x}}
    \foreach \x in {7,...,9}{\circle{-13.5+\x}{-0.5+\x}}
    \foreach \x in {2,...,3}{\circle{0.5+\x}{-0.5+\x}}
    \foreach \x in {4,...,10}{\circle{7.5-\x}{-0.5+\x}}
    \flatRD{-0.5}{0.5}
    \flatRD{0.5}{1.5}
    \flatLD{1.5}{1.5}
    \flatRD{-3.5}{9.5}
    \draw[semithick,decorate,decoration={brace}] (-6.65,6.65) -- (-3.65,9.65) node[midway,xshift=-5pt,yshift=5pt] {$x$};
    \draw[semithick,decorate,decoration={brace}] (-2.35,9.65) -- (3.65,3.65) node[midway,xshift=5pt,yshift=5pt] {$y$};
    \draw[semithick,decorate,decoration={brace}] (0.65,1.35) -- (-0.35,0.35) node[midway,xshift=5pt,yshift=-5pt] {$k$};
    \draw[semithick,decorate,decoration={brace}] (3.65,2.35) -- (2.65,1.35) node[midway,xshift=9pt,yshift=-5pt] {$x\!-\!k$};
  \end{tikzpicture},
\end{equation}
where $0\leq k \leq x$. Note that 
\be
\label{eq:Cflatwone}
w_1(t-x+1,t+x) = \expval{C_{\SflatSA\SflatSA}(x,t)}_{\rpc}. 
\ee 
Next, we observe that $w_k(x,y)$ and $v_k(x,y)$ fulfil a closed system of recurrence relations given by 
\begin{equation}
\begin{aligned}
\label{eq:recOriginal}
  w_k(x,y)&=\frac{\sqrt{q}}{1+q}
  \left(w_{k-1}(x-1,y)+v_k(x,y)\right),\\
  v_k(x,y)&=\frac{\sqrt{q}}{1+q}
  \left(w_k(x,y-1)+v_{k+1}(x,y)\right),
  \end{aligned}
\end{equation}
with boundary conditions
\begin{equation}
  \begin{aligned}
  \label{eq:recOriginalBC}
    w_0(x,y)&=\frac{1}{\sqrt{q}},    \qquad
    w_1(1,1)=\frac{2}{(1+q)},   \\
    v_0(0,1)&=1, \qquad
    v_1(1,y)=w_1(1,y-1),        \\
    v_0(x,1)&=\frac{1}{1+q}\left(1+v_0(x-1,1)\right), \\
    v_{x}(x,y)&=\frac{1}{\sqrt{q}} w_x(x,y-1),\\
    v_k(x,1)&=\frac{1}{\sqrt{q}} v_{k-1}(x-1,1)=
    {q}^{-k/2} v_{0}(x-k,1).
  \end{aligned}
\end{equation}
The validity of Eqs.~\eqref{eq:recOriginal} and \eqref{eq:recOriginalBC} can be straightforwardly proven diagrammatically, by repeated application of the graphical identities in Eq.~\eqref{eq:WdEval2}. For example, considering \eqref{eq:wk} and repeatedly applying the second line of \eqref{eq:WdEval2} starting from the bottom corner one readily obtains the first of \eqref{eq:recOriginal}.

Remarkably Eqs.~\eqref{eq:recOriginal} can be solved explicitly for any boundary condition by writing a generating function fulfilling a two-variable recurrence relation and solving it via the so called kernel method~\cite{pemantle2013analytic}. The explicit derivation is presented in App.~\ref{app:recSol}, while the relevant result in our case reads 
\begin{equation}
 \label{eq:w1solution}
  \begin{aligned}
    w_1(x,&y)=\frac{1}{q}+
  \frac{q^2-1}{q}\left(\frac{\sqrt{q}}{1+q}\right)^{2x+2y}\\
    &\cross[A_{x,y}(\alpha)A_{y-1,x}(\alpha)-A_{x,y-1}(\alpha)A_{y,x}(\alpha)],
  \end{aligned}
\end{equation}
where $\alpha$ and $A_{x,y}(z)$ are respectively defined as
\be
\alpha=1+\frac{1}{q}, 
\label{eq:alpha}
\ee
and
\begin{equation}\label{eq:defAky}
    A_{x,y}(z)=\sum_{t=1}^{y}\binom{x-1+y-t}{x-1} z^t+\delta_{y,0}\delta_{x,0}.
\end{equation}
Substituting back into Eq.~\eqref{eq:Cflatwone} and then in \eqref{eq:CgenCflat} we finally obtain 
\begin{widetext}
\begin{equation}
  \expval{C_{\mu\nu}(x,t)}_{\rpc} = \frac{o_{\mu} o_{\nu}}{1-q^{-2}} \left[\frac{\sqrt{q}}{1+q}\right]^{4t} (A_{t-x+1,t+x}(\alpha)A_{t+x-1,t-x+1}(\alpha)-A_{t-x+1,t+x-1}(\alpha)A_{t+x,t-x+1}(\alpha)).
  \label{eq:avecorrresult}
\end{equation}
\end{widetext}

Interestingly, the functional form of the correlation Eq.~\eqref{eq:avecorrresult} is exactly the same as that of averaged correlations \emph{squared} in random unitary circuits, i.e.\ $\expval*{C_{\mu\nu}(x,t)^2}_{\ruc}$. 
This can be seen by noting that also $\expval*{C_{\mu\nu}(x,t)^2}_{\ruc}$ fulfils the system of equations \eqref{eq:recOriginal} up to a simple redefinition of the parameters. We begin by considering the diagrammatic representation of the averaged squared correlation in random unitary circuits
\begin{equation} \label{eq:averaged_Cab_squared}
  \mkern-4mu
  \expval*{C^2_{\mu\nu}(x,t)}_{\ruc}=
  \mkern-25mu
  \begin{tikzpicture}[baseline={([yshift=-2.2ex]current bounding box.center)},scale=0.5]
    \prop{0}{0}{colIHaar}{4}
    \foreach \x in {-1,1}{\prop{\x}{1}{colIHaar}{4}}
    \foreach \x in {-2,0,2}{\prop{\x}{2}{colIHaar}{4}}
    \foreach \x in {-3,-1,...,3}{\prop{\x}{3}{colIHaar}{4}}
    \foreach \x in {-4,-2,0,2}{\prop{\x}{4}{colIHaar}{4}}
    \foreach \x in {-5,-3,-1,1}{\prop{\x}{5}{colIHaar}{4}}
    \foreach \x in {-6,-4,-2,0}{\prop{\x}{6}{colIHaar}{4}}
    \foreach \x in {-5,-3,-1}{\prop{\x}{7}{colIHaar}{4}}
    \foreach \x in {-4,-2}{\prop{\x}{8}{colIHaar}{4}}
    \foreach \x in {-3}{\prop{\x}{9}{colIHaar}{4}}
    \foreach \x in {0,...,6}{\circle{-0.5-\x}{-0.5+\x}}
    \foreach \x in {7,...,9}{\circle{-13.5+\x}{-0.5+\x}}
    \foreach \x in {1,...,3}{\circle{0.5+\x}{-0.5+\x}}
    \foreach \x in {4,...,10}{\circle{7.5-\x}{-0.5+\x}}
    \blackcircle{0.5}{-0.5}
    \blackcircle{-3.5}{9.5}
    \node at (0.9,-0.9) {\scalebox{0.9}{$\mathcal{O}_{\mu}$}};
    \node at (-3.1,9.9) {\scalebox{0.9}{$\mathcal{O}_{\nu}$}};
    \draw[black,|<->|] (-3.5,-1.5) -- (0.5,-1.5) node[midway,above] {$2x$};
    \draw[black,|<->|] (4,-0.5) -- (4,9.5) node[pos=0.75,left] {$2t$};
  \end{tikzpicture}.
  \mkern-4mu
\end{equation}
Here the dark circle represents
\begin{equation}
  \ket*{\blackcircleSA{\mu}^{(4)}} = (\mathcal{O}_{\mu}\otimes\1)^{\otimes 2}
  \ket*{\circleSA^{(4)}} 
  = \begin{tikzpicture}[baseline={([yshift=-0.6ex]current bounding box.center)},scale=0.75]
    \gridLine{0.5}{-0.5}{0.5}{0.25}
    \blackcircle{0.5}{-0.5}
    \node at (1,-0.5) {\scalebox{0.9}{$\mathcal{O}_{\mu}$}};
  \end{tikzpicture}  \,,
\end{equation}
which can be straightforwardly expanded in terms of  $\ket*{\circleSA^{(4)}}$ and $\ket*{\squareSA^{(4)}}$. This leads to
\begin{equation} \label{eq:CgenCsquare}
  \expval{C^2_{\mu\nu}(x,t)}_{\ruc}=
  \frac{\expval*{C^2_{\scalebox{0.6}{\!\!\squareSA\!\!\!\squareSA\!\!}}(x,t)}_{\ruc}-\frac{1}{q^2}}
  {\left(1-\frac{1}{q^2}\right)^2}
\end{equation}
which is the direct analogue of Eq.~\eqref{eq:CgenCflat}. In fact, the numerical coefficients of Eq.~\eqref{eq:CgenCsquare} can be obtained from those of Eq.~\eqref{eq:CgenCflat} by performing the mapping 
\be
o_\mu\mapsto1, \qquad q\mapsto q^2\,. 
\label{eq:mapping}
\ee 
Finally, to evaluate $\expval*{C^2_{\scalebox{0.6}{\!\!\squareSA\!\!\!\squareSA\!\!}}(x,t)}_{\ruc}$, we recall the exact equivalence between \eqref{eq:UdEval2} and \eqref{eq:WdEval2}, which implies that one can directly repeat the manipulations performed in the first part of this subsection with the only (trivial) difference arising from the different numerical factor on the r.h.s.\ of the equations on second line of~\eqref{eq:UdEval2} and~\eqref{eq:WdEval2}, which is accounted for by the second of Eq.~\eqref{eq:mapping}. This leads to the promised correspondence. 

\subsection{Out-of-time-ordered correlation functions}
\label{sec:OTOCs}
To provide a quantitative characterisation of operator spreading we consider OTOCs. In particular, we compute the \emph{averaged} OTOCs between two traceless Hermitian one-site operators $\mathcal{O}_{\mu}$, and $\mathcal{O}_{\nu}$, as defined in Eq.~\eqref{eq:defOTOCs}, the r.h.s.\ of which can be conveniently split into two contributions 
\begin{equation}\label{eq:defOTOCs2}
\begin{aligned}
  \expval{{O}_{\mu\nu}(x,t)}_{\rpc}
  =
  &
  \overbrace{
    \frac{\expval*{\tr[\smash{\mathcal{O}_{\mu}^2(0,0)\mathcal{O}_{\nu}^2(x,t)}]}_{\rpc}}
    {\tr\1}}^{\tilde{O}_{\mu\nu}^{(1)}(x,t)}
  \\
  &-
  \underbrace{
    \frac{\expval*{\tr[\smash{(\mathcal{O}_{\mu}(0,0)\mathcal{O}_{\nu}(x,t))^2}]}_{\rpc}}
    {\tr\1}}_{\tilde{O}_{\mu\nu}^{(2)}(x,t)}.
\end{aligned}
\end{equation}
We first note that the first term is given as
\begin{equation}
  \tilde{O}_{\mu\nu}^{(1)}(x,t)=
    1+
    \tilde{o}_{\mu}\tilde{o}_{\nu}\left(q \expval{C_{\SflatSA\SflatSA}(x,t)}_{\rpc}-1\right),
\end{equation}
where $\expval{C_{\SflatSA\SflatSA}(x,t)}_{\rpc}$ is the correlation function appearing in Eq.~\eqref{eq:CgenCflat}, we used $\tr[\mathcal{O}_{\mu}^2]=q$ and we introduced 
\begin{equation}
    \tilde{o}_{\mu}
    = \frac{1}{q-1}\left(\frac{1}{q} \sum_{i,j=0}^{q-1} \mel{i}{\mathcal{O}_{\mu}^2}{j}-1\right).
\end{equation}
Note that if the operator $\mathcal{O}_{\mu}$ is \emph{diagonal} in the computational basis we have $\tilde{o}_{\mu}=0$. Interestingly, this also holds for randomly selected Hermitian operators with Gaussian-distributed matrix ensembles.

The second term on the r.h.s.\ of Eq.~\eqref{eq:defOTOCs2} is given by the following diagram
\begin{equation}\label{eq:relDiagramOTOCs}
  \tilde{O}_{\mu\nu}^{(2)}(x,t)
  = q^{2t+1} \mkern-24mu
  \begin{tikzpicture}[baseline={([yshift=-2.2ex]current bounding box.center)},scale=0.5]
    \prop{0}{0}{colPerm}{4}
    \foreach \x in {-1,1}{\prop{\x}{1}{colPerm}{4}}
    \foreach \x in {-2,0,2}{\prop{\x}{2}{colPerm}{4}}
    \foreach \x in {-3,-1,...,3}{\prop{\x}{3}{colPerm}{4}}
    \foreach \x in {-4,-2,0,2}{\prop{\x}{4}{colPerm}{4}}
    \foreach \x in {-5,-3,-1,1}{\prop{\x}{5}{colPerm}{4}}
    \foreach \x in {-6,-4,-2,0}{\prop{\x}{6}{colPerm}{4}}
    \foreach \x in {-5,-3,-1}{\prop{\x}{7}{colPerm}{4}}
    \foreach \x in {-4,-2}{\prop{\x}{8}{colPerm}{4}}
    \foreach \x in {-3}{\prop{\x}{9}{colPerm}{4}}
    \foreach \x in {0,...,6}{\squareD{-0.5-\x}{-0.5+\x}}
    \foreach \x in {7,...,9}{\circle{-13.5+\x}{-0.5+\x}}
    \foreach \x in {1,...,3}{\squareD{0.5+\x}{-0.5+\x}}
    \foreach \x in {4,...,10}{\circle{7.5-\x}{-0.5+\x}}
    \blacksquareD{0.5}{-0.5}
    \blackcircle{-3.5}{9.5}
    \node at (0.8,-1) {\scalebox{0.9}{$\mathcal{O}_{\nu}$}};
    \node at (-3.2,10) {\scalebox{0.9}{$\mathcal{O}_{\mu}$}};
    \draw[semithick,decorate,decoration={brace}] (-6.65,6.65) -- (-3.65,9.65) node[midway,xshift=-5pt,yshift=5pt,rotate=45] {$t-x+1$};
    \draw[semithick,decorate,decoration={brace}] (-2.35,9.65) -- (3.65,3.65) node[midway,xshift=5pt,yshift=5pt,rotate=-45] {$t+x$};
  \end{tikzpicture},
\end{equation}
where, in analogy with Eq.~\eqref{eq:blackcirclestate}, we introduced
\begin{equation}\label{eq:blackcirclesquare4}
  \begin{aligned}
    \ket*{\blackcircleSA{\mu}^{(4)}} &= 
    (\mathcal{O}_{\mu}\otimes\1)^{\otimes 2} \ket*{\circleSA^{(4)}} 
    = \begin{tikzpicture}[baseline={([yshift=-0.6ex]current bounding box.center)},scale=0.75]
      \gridLine{0.5}{-0.5}{0.5}{0.25}
      \blackcircle{0.5}{-0.5}
      \node at (1,-0.5) {\scalebox{0.9}{$\mathcal{O}_{\mu}$}};
    \end{tikzpicture},\\
    \ket*{\blacksquareSA{\mu}^{(4)}} &= 
    (\mathcal{O}_{\mu}\otimes\1)^{\otimes 2} \ket*{\squareSA^{(4)}} 
    = \begin{tikzpicture}[baseline={([yshift=-0.6ex]current bounding box.center)},scale=0.75]
      \gridLine{0.5}{-0.5}{0.5}{0.25}
      \blacksquare{0.5}{-0.5}
      \node at (1,-0.5) {\scalebox{0.9}{$\mathcal{O}_{\mu}$}};
    \end{tikzpicture}  \,. 
  \end{aligned}
\end{equation}
For generic observables the states $\ket*{\blackcircleSA{\mu}^{(4)}}$ and $\ket*{\blacksquareSA{\nu}^{(4)}}$ have non-zero overlap with all of the relevant states $\ket*{\pi_j^{(4)}}$, which makes the exact contraction of the above diagram unfeasible. We are, however, able to treat exactly the case where at least one of the operators is \emph{diagonal} in the computational basis. This follows from the fact that for a diagonal operator $\mathcal{O}_{\mathrm{d}}$, the projection of $\ket*{\blackcircleSA{\rm d}^{(4)}}$ to the invariant states includes only $\ket*{\circleSA^{(4)}}$ and a single additional state (cf.\ App.~\ref{app:partitions})
\begin{equation} \label{eq:defComb4}
  \ket*{\combSA^{(4)}}:=\frac{1}{\sqrt{q}} \ket*{\pi_1^{(4)}} = 
  \begin{tikzpicture}[baseline={([yshift=-0.6ex]current bounding box.center)},scale=0.75]
    \gridLine{-0.3}{-0.3}{-0.3}{0.4}
    \comb{-0.3}{-0.3}
  \end{tikzpicture},
\end{equation}
while we also have
\begin{align}
  \label{eq:otocLocalRelations}
  \begin{gathered}
    \begin{tikzpicture}[baseline={([yshift=-0.6ex]current bounding box.center)},scale=0.75]
      \prop{0}{0}{colPerm}{4}
      \circle{-0.5}{0.5}
      \circle{0.5}{0.5}
    \end{tikzpicture}=
    \begin{tikzpicture}[baseline={([yshift=-0.6ex]current bounding box.center)},scale=0.75]
      \gridLine{-0.3}{-0.3}{-0.3}{0.45}
      \gridLine{0.3}{-0.3}{0.3}{0.45}
      \circle{-0.3}{0.45}
      \circle{0.3}{0.45}
    \end{tikzpicture},\qquad
    \begin{tikzpicture}[baseline={([yshift=-0.6ex]current bounding box.center)},scale=0.75]
      \prop{0}{0}{colPerm}{4}
      \combD{-0.5}{0.5}
      \combD{0.5}{0.5}
    \end{tikzpicture}=
    \begin{tikzpicture}[baseline={([yshift=-0.6ex]current bounding box.center)},scale=0.75]
      \gridLine{-0.3}{-0.3}{-0.3}{0.45}
      \gridLine{0.3}{-0.3}{0.3}{0.45}
      \comb{-0.3}{0.45}
      \comb{0.3}{0.45}
    \end{tikzpicture},\\
    \begin{tikzpicture}[baseline={([yshift=-0.6ex]current bounding box.center)},scale=0.75]
      \prop{0}{0}{colPerm}{4}
      \circle{-0.5}{0.5}
      \combD{0.5}{0.5}
    \end{tikzpicture}=
    \begin{tikzpicture}[baseline={([yshift=-0.6ex]current bounding box.center)},scale=0.75]
      \prop{0}{0}{colPerm}{4}
      \combD{-0.5}{0.5}
      \circle{0.5}{0.5}
    \end{tikzpicture}=
    \frac{\sqrt{q}}{q+1}
    \left(
    \begin{tikzpicture}[baseline={([yshift=-0.6ex]current bounding box.center)},scale=0.75]
      \gridLine{-0.3}{-0.3}{-0.3}{0.45}
      \gridLine{0.3}{-0.3}{0.3}{0.45}
      \circle{-0.3}{0.45}
      \circle{0.3}{0.45}
    \end{tikzpicture}
    +\begin{tikzpicture}[baseline={([yshift=-0.6ex]current bounding box.center)},scale=0.75]
      \gridLine{-0.3}{-0.3}{-0.3}{0.45}
      \gridLine{0.3}{-0.3}{0.3}{0.45}
      \comb{-0.3}{0.45}
      \comb{0.3}{0.45}
    \end{tikzpicture}\right).
  \end{gathered}
\end{align}
Analogously to the treatment of two-point correlation functions, this allows one to formulate a difference equation whose solution gives the diagram in Eq.~\eqref{eq:relDiagramOTOCs}. In fact, as it is shown in App.~\ref{sec:DiagonalOTOCs}, the resulting system of recurrence relations can be exactly mapped into Eqs.~\eqref{eq:recOriginal} and \eqref{eq:recOriginalBC} upon redefinition of the parameters. This leads to the exact expression for the OTOC between a diagonal observable $\mathcal{O}_{\mathrm{d}}$ and an arbitrary observable $\mathcal{O}_{\nu}$ given in Eq.~\eqref{eq:OTOCrandomperm}, namely using the shorthand notation
\begin{equation}
    \alpha=1+q,\qquad o_{\mu,2}=\frac{1}{q}\sum_{i=0}^{q-1} \mel{i}{\mathcal{O}_{\mu}}{i},
\end{equation}
the averaged OTOC takes the following form,
\begin{widetext}
  \begin{equation}
    \label{eq:OTOCRP}
    \expval{{O}_{d\nu}(x,t)}_{\rpc} 
    = \frac{q (1-o_{\nu,2})}{q^2-1} \left[\frac{\sqrt{q}}{1+q}\right]^{4t}
    (A_{t-x+1,t+x}(\alpha)A_{t+x-1,t-x+1}(\alpha)-A_{t-x+1,t+x-1}(\alpha)A_{t+x,t-x+1}(\alpha)),
  \end{equation}
\end{widetext}
cf.\ Eq.~\eqref{eq:defAky} for the definition of $A_{x,y}(z)$.

Interestingly, also in this case we have an exact correspondence between the result found for random permutations and that of random unitary circuits. In this case, the precise statement is that the OTOC between a diagonal and a generic operator averaged over random permutations, i.e.\ Eq.~\eqref{eq:OTOCrandomperm}, coincides exactly with the OTOC among two local operators averaged over random unitary circuits (cf.\ Refs.~\cite{nahum2018operator, bertini2020scrambling}), if one performs the mapping (cf.\ Eq.~\eqref{eq:o2mu})
\begin{equation}
o_{\mu,2}\mapsto 0, \qquad q\mapsto q^2\,. 
\label{eq:mapping2}
\end{equation}
To show this we start from the expression of the averaged OTOC in random unitary circuits in terms of the solution of a system of recurrence relations given in Ref.~\cite{bertini2020scrambling}. The latter is derived proceeding along the lines of the previous subsection: one writes a diagrammatic representation for the object of interest and repeatedly applies Eq.~\eqref{eq:UdEval2}. In our notation the result can be expressed as 
\begin{equation} \label{eq:OTOCRUCinitial}
  \mkern-8mu
  \expval{{O}_{\mu\nu}(x,t)}_{\ruc} \!=\! \frac{q^4}{(q^2-1)^2} \left(1\!-\!q^{2t-1} a_{t-x}(1,t+x)\right),
  \mkern-8mu
\end{equation}
where $a_k(x,y)$ fulfils
\begin{equation}
\begin{aligned}
\label{eq:OTOCRUCsys}
  a_k(x,y)&=\frac{1}{1+q^2} a_{k}(x-1,y)+\frac{q}{1+q^2} b_{k}(x,y),\\
  b_k(x,y)&=\frac{1}{1+q^2} a_{k}(x,y-1)+\frac{q}{1+q^2} b_{k-1}(x,y),
  \end{aligned}
\end{equation}
with boundary conditions 
\begin{equation}
  \begin{aligned}
  \label{eq:OTOCBC}
    a_k(x,0)&=\frac{1}{q^{|k-1|}},    \qquad
    a_k(k,y)&=\frac{1}{q^{|y+k-1|}},   \\
    b_{0}(x,y)&=a_{0}(x,y).
  \end{aligned}
\end{equation}
As shown in App.~\ref{sec:OTOCsRUC} this system can again be brought to the same form as Eqs.~\eqref{eq:recOriginal} and \eqref{eq:recOriginalBC} and explicitly solved. This yields precisely the expression obtained from Eq.~\eqref{eq:OTOCRP} via the mapping in Eq.~\eqref{eq:mapping2}.

\subsection{Entanglement Growth after a Quench}
\label{sec:entanglement}

As a further characterisation of quantum dynamics we study the rate of growth of entanglement after a quantum quench, which we conveniently measure evaluating the \emph{averaged purity}. Specifically we initialise the system in a random product state 
\begin{equation}
  \ket{\psi(0)}=\bigotimes_{j=1}^{2L} \ket*{\psi_j},
\end{equation}
where $\ket*{\psi_j}$ are independently distributed single-qudit random states~\footnote{The latter can be obtained by considering a given normalised state, say $\ket{0}$, and applying a random unitary matrix drawn according to the Haar distribution over $U(q)$.}. Then we let the system evolve according to Eq.~\eqref{eq:defTE}, until, at time $t< 2L$, we compute the purity with respect to a bipartition of the system in two halves, and then average it over all possible choices of gates and initial states
\begin{equation}
 \expval{P(t)}=\expval{\tr[\left(\tr_{\mathbb{Z}_L/2}\ketbra{\psi(t)}{\psi(t)}\right)^2]}.
\end{equation}
Considering open boundary conditions (i.e.\ only one boundary between the two subsystems) we find that the above quantity can be represented diagrammatically as follows 
\begin{equation}
\label{eq:P}
  \expval{P(t)} =
  \begin{tikzpicture}[baseline={([yshift=-0.6ex]current bounding box.center)},scale=0.5]
    \prop{1}{1}{white}{4}
    \foreach \x in {0,2}{\prop{\x}{0}{white}{4}}
    \foreach \x in {-1,1,3}{\prop{\x}{-1}{white}{4}}
    \foreach \x in {-2,0,...,4}{\prop{\x}{-2}{white}{4}}
    \foreach \x in {-3,-1,...,5}{\prop{\x}{-3}{white}{4}}
    \foreach \x in {-4,-2,...,6}{\prop{\x}{-4}{white}{4}}
    \foreach \x in {-5,...,0}{\circle{\x+0.5}{\x+1.5}}
    \foreach \x in {-5,...,0}{\squareD{-\x+1.5}{\x+1.5}}
    \foreach \x in {-5,...,6}{\inState{\x+0.5}{-4.5}}
    \draw[semithick,decorate,decoration={brace}] (1.5,1.85) -- (6.85,-3.5) node[midway,xshift=7.5pt,yshift=7.5pt] {$2t$};
  \end{tikzpicture},
\end{equation}
where we introduced the following graphical representation of the \emph{averaged} initial state
  \begin{equation}
 \begin{tikzpicture}[baseline={([yshift=-0.6ex]current bounding box.center)},scale=0.5]
   \gridLine{0}{0}{0}{0.75}
   \inState{0}{0}
 \end{tikzpicture}=
  q\expval*{\left(\ket{\psi}\otimes\ket{\psi}^{\ast}\right)^{\otimes 2}}:=
  \ket*{\inStateSA}.
\end{equation}
Note that all the gates outside a light cone emanating from the subsystem's boundary can be removed by using the unitarity of the local gates. We also emphasise that in the above equation we did not specify over what ensemble the average is performed and, therefore, we did not colour the averaged gates.  

When considering the average over random unitary gates the diagram in Eq.~\eqref{eq:P} can be directly contracted. Indeed, one can just repeatedly apply the relations on the second line of \eqref{eq:UdEval2} starting from the top corner of the triangle. The result reads~\cite{nahum2018operator}
\be
\label{eq:Purityru}
\expval{P(t)}_{\ruc}= \left(\frac{2q}{q^2+1}\right)^{2t}. 
\ee
Instead, in the case of random permutations the situation is more complicated. This is because the diagram at the top corner of the triangle, i.e.\  
\be
\mathcal{Q}^{(4)}\ket*{\circleSA\,\squareSA},
\ee
produces a linear combination of all the $15$ states in Eq.~\eqref{eq:doublepart} for $n=4$. This generates a rapidly growing number of terms and makes it very difficult to attain a close-form expression. To simplify the treatment we evaluate the expressions in the in the large-$q$ limit: this introduces a substantial simplification because, at leading order, we only need to account for $\ket*{\circleSA^{(4)}}$, $\ket{\squareSA^{(4)}}$, and $\ket{\combSA^{(4)}}$ (cf.~\eqref{eq:defComb4}). Indeed, at leading order we have the following relevant relations
\begin{equation}\label{eq:purityLocalRelations}
  \begin{aligned}
    \begin{tikzpicture}[baseline={([yshift=-0.6ex]current bounding box.center)},scale=0.75]
      \prop{0}{0}{colPerm}{4}
      \circle{-0.5}{0.5}
      \squareD{0.5}{0.5}
    \end{tikzpicture}&=
    \frac{1}{q}\left(
    \begin{tikzpicture}[baseline={([yshift=-0.6ex]current bounding box.center)},scale=0.75]
      \gridLine{-0.3}{-0.3}{-0.3}{0.45}
      \gridLine{0.3}{-0.3}{0.3}{0.45}
      \circle{-0.3}{0.45}
      \circle{0.3}{0.45}
    \end{tikzpicture}
    +\begin{tikzpicture}[baseline={([yshift=-0.6ex]current bounding box.center)},scale=0.75]
      \gridLine{-0.3}{-0.3}{-0.3}{0.45}
      \gridLine{0.3}{-0.3}{0.3}{0.45}
      \square{-0.3}{0.45}
      \square{0.3}{0.45}
    \end{tikzpicture}
    +\begin{tikzpicture}[baseline={([yshift=-0.6ex]current bounding box.center)},scale=0.75]
      \gridLine{-0.3}{-0.3}{-0.3}{0.45}
      \gridLine{0.3}{-0.3}{0.3}{0.45}
      \comb{-0.3}{0.45}
      \comb{0.3}{0.45}
    \end{tikzpicture}
    \right),\\
    \begin{tikzpicture}[baseline={([yshift=-0.6ex]current bounding box.center)},scale=0.75]
      \prop{0}{0}{colPerm}{4}
      \circle{-0.5}{0.5}
      \combD{0.5}{0.5}
    \end{tikzpicture}&=
    (q-1)q^{-\frac{3}{2}}
    \left(
    \begin{tikzpicture}[baseline={([yshift=-0.6ex]current bounding box.center)},scale=0.75]
      \gridLine{-0.3}{-0.3}{-0.3}{0.45}
      \gridLine{0.3}{-0.3}{0.3}{0.45}
      \circle{-0.3}{0.45}
      \circle{0.3}{0.45}
    \end{tikzpicture}
    +\begin{tikzpicture}[baseline={([yshift=-0.6ex]current bounding box.center)},scale=0.75]
      \gridLine{-0.3}{-0.3}{-0.3}{0.45}
      \gridLine{0.3}{-0.3}{0.3}{0.45}
      \comb{-0.3}{0.45}
      \comb{0.3}{0.45}
    \end{tikzpicture}\right),\\
    \begin{tikzpicture}[baseline={([yshift=-0.6ex]current bounding box.center)},scale=0.75]
      \prop{0}{0}{colPerm}{4}
      \combD{-0.5}{0.5}
      \squareD{0.5}{0.5}
    \end{tikzpicture}&=
    (q-1)q^{-\frac{3}{2}}
    \left(
    \begin{tikzpicture}[baseline={([yshift=-0.6ex]current bounding box.center)},scale=0.75]
      \gridLine{-0.3}{-0.3}{-0.3}{0.45}
      \gridLine{0.3}{-0.3}{0.3}{0.45}
      \square{-0.3}{0.45}
      \square{0.3}{0.45}
    \end{tikzpicture}
    +\begin{tikzpicture}[baseline={([yshift=-0.6ex]current bounding box.center)},scale=0.75]
      \gridLine{-0.3}{-0.3}{-0.3}{0.45}
      \gridLine{0.3}{-0.3}{0.3}{0.45}
      \comb{-0.3}{0.45}
      \comb{0.3}{0.45}
    \end{tikzpicture}\right).
  \end{aligned}
\end{equation}
To find a recurrence relation for $\expval{P(t)}_{\rpc}$ it is useful to introduce an auxiliary object defined as 
\begin{equation} \label{eq:Q}
  \mkern-4mu
  Q(m,n)=\mkern-30mu
  \begin{tikzpicture}[baseline={([yshift=-0.6ex]current bounding box.center)},scale=0.5]
    \foreach \x in {0,2,4}{\prop{\x}{0}{colPerm}{4}}
    \foreach \x in {-1,1,...,5}{\prop{\x}{-1}{colPerm}{4}}
    \foreach \x in {-2,0,...,6}{\prop{\x}{-2}{colPerm}{4}}
    \foreach \x in {-3,-1,...,7}{\prop{\x}{-3}{colPerm}{4}}
    \foreach \x in {-4,-2,...,8}{\prop{\x}{-4}{colPerm}{4}}
    \foreach \x in {-5,...,-1}{\circle{\x+0.5}{\x+1.5}}
    \foreach \x in {-5,...,-1}{\squareD{4-\x-0.5}{\x+1.5}}
    \foreach \x in {-5,...,8}{\inState{\x+0.5}{-4.5}}
    \foreach \x in {0,...,3}{\combD{\x+0.5}{0.5}}
    \draw[semithick,decorate,decoration={brace}] 
    (4.5,0.85)--(8.85,-3.5) node[midway,xshift=7.5pt,yshift=7.5pt] {$m$};
    \draw[semithick,decorate,decoration={brace}] (0.35,0.85)--(3.85,0.85) node[midway,above] {$n$};
  \end{tikzpicture}\mkern-15mu,
  \mkern-4mu
\end{equation}
where $n\ge0$ is assumed to be even. In terms of this diagram the averaged purity is
\begin{equation}
  \expval{P(t)}_{\rpc}=Q(2t,0).
\end{equation}
Considering~\eqref{eq:Q} and repeatedly using the local relations \eqref{eq:purityLocalRelations} we obtain the following recursive relation  
\begin{equation}
  \mkern-4mu
  \begin{aligned}
    Q_{2}(m,n)&= q^{-1}\big(2 Q_2(m-1,n)+Q_2(m-1,n-2)\\
    &+Q_2(m-1,n+2)\big),\\
    Q_2(m,0)&= q^{-1}\left(2 Q_2(m-1,0)+ Q_2(m-1,2)\right),\\
    Q_2(0,n)&= \braket*{\combSA}{\inStateSA}^n.
  \end{aligned}
  \mkern-4mu
\end{equation}
The solution of this relation can be found by elementary methods and expressed as 
\begin{equation}
\begin{aligned}
  Q_2(m,2n)&=q^{-m}\sum_{k=0}^{m+n}\braket*{\combSA}{\inStateSA}^{2k} \binom{2m}{m+n-k}\\
  &-q^{-m}\sum_{k=0}^{m+n}\braket*{\combSA}{\inStateSA}^{2k} \binom{2m}{m-2-n-k}.
\end{aligned}
\end{equation}
This gives
\begin{equation}\label{eq:solPurity}
  \mkern-8mu
  \expval{P(t)}_{\rpc} \mkern-5mu=\mkern-8mu 
  \sum_{k=0}^{2t}\!\left[\!\binom{4t}{2t-k}\!-\!\binom{4t}{2t-k-2}\!\right]
  \!\frac{\braket*{\combSA}{\inStateSA}^{2k}}{q^{2t}}.
  \mkern-8mu
\end{equation}
We emphasise that this solution accurately describes the averaged purity \emph{only} at the
leading-order in $q^{-1}$. To find it, we note that the averaged random initial state projected to invariant states becomes
\begin{equation}
  \ket*{\inStateSA}\to \frac{1}{1+\frac{1}{q}}
  \left(\ket*{\circleSA}+\ket*{\squareSA}\right),
\end{equation}
which gives 
\begin{equation}
  \braket*{\combSA}{\inStateSA}=\frac{2}{(1+\frac{1}{q})\sqrt{q}}\approx\frac{2}{\sqrt{q}}.
\end{equation}
This means that the sum in Eq.~\eqref{eq:solPurity} gives its leading contribution for $k=0$ and we have
\begin{equation}
\begin{aligned}
 \expval{P(t)}_{\rpc} &= q^{-2t}\left[\binom{4t}{2t}-\binom{4t}{2t-2}\right]\\
  &=q^{-2t}\frac{2 (4t+1)!}{(2t+2)! (2t)!} \simeq \left(\frac{4}{q}\right)^{2t},
  \end{aligned}
\end{equation}
where in the last step we used Stirling's formula. Comparing with the large $q$ limit of Eq.~\eqref{eq:Purityru} we see that for random permutations the purity is enhanced by a factor $2^{2t}$ compared to random unitaries. This means that random permutation create less entanglement that random unitary, however, they produce the same qualitative behaviour.  

\subsubsection{Special product states}
Even though we focussed on random product states our approach can be directly applied to all initial product states that become translational invariant upon the application of a layer of averaged gates. 

An interesting example of states with this property are product states in the computational basis. Since the permutation gates map these states into other computational-basis states, the time-evolved state is always in a product form, and there is no entanglement growth. This is consistent with the solution in Eq.~\eqref{eq:solPurity}. Indeed, in this case we have
\begin{equation}
  \braket*{\combSA}{\inStateSA}\to\sqrt{q},
\end{equation}
and therefore
\begin{equation}
  \expval{P(t)}_{\rpc} 
  =\binom{4t}{0}-\binom{4t}{-2}=1.
\end{equation}

Another example of state treatable with our approach, and that leads to no entanglement growth, is the uniform superposition of all computation-basis states 
\begin{equation}
  \ket{\psi(0)}=\left(\frac{1}{\sqrt{q}} \sum_{i=0}^{q-1}\ket{i}\right)^{\otimes 2L}.
\end{equation}
Since permutation gates map each computational basis state in a single computational basis state this flat superposition is left invariant by the time evolution, and its purity is pinned to one. As we show in App.~\ref{app:purityFlat}, however, for this state one cannot directly apply the large $q$ formula in Eq.~\eqref{eq:solPurity} because the simplified local relations in Eq.~\eqref{eq:purityLocalRelations} do not give the leading contribution.     

\section{Classical Dynamics}
\label{sec:class}
In this section we characterise the classical dynamics generated by random permutation gates.  As discussed in Sec.~\ref{sec:setting}, in this setting one can introduce suitable probes of many-body dynamics that show the same qualitative (and sometimes quantitative) behaviours as the standard probes used in the quantum setting. 

\subsection{Decorrelators and damage spreading}
\label{sec:class1}
While averaged expectation values and correlation functions are immediately zero in the classical setting, we can consider the decorrelator introduced in Sec.~\ref{sec:resultsclassical}. In essence, this quantity measures how a local perturbation spreads under classical dynamics. 
It is defined by taking two copies of the initial state, changing the state of one of the copies at $(t=0, x=0)$ by applying a one-site traceless probability-conserving operator $\mathcal{O}_{\alpha}$ (e.g.\ a bit flip for $q=2$), and letting both systems undergo the same deterministic time evolution. The decorrelator measures the probability that the two configurations disagree at position $x$ and time $t$ upon taking a uniform average over all initial configurations (and random permutation gates). Explicitly, the decorrelator can be expressed as
\begin{equation}
  D(x,t)=1-\frac{1}{q^{2L}}\tr[\mathcal O_{f}(x) \mathcal O_{\alpha}(0,t)].
\end{equation}
Here $\mathcal{O}_{\alpha}$, $\mathcal{O}_{\alpha}(x,t)$ denotes the operator
$\mathcal{O}_{\alpha}$ applied at position $x$ and time-evolved for $t$ time
steps, and $\mathcal O_f(x)$ is the operator that forces the two copies to be
the same at the site $x$
\begin{equation}
  \frac{\mathcal O_f(x)}{q^{2L-1}} = 
  \ketbra*{\flatSA}^{\otimes(2x-1)}\otimes\1\otimes \ketbra*{\flatSA} ^{\otimes(2L-2x)},
\end{equation}
where $\ket{\flatSA}$ is a shorthand for
\begin{equation}\label{eq:defFlatClass}
  \ket{\flatSA}:=\ket*{\flatSA^{(1)}}=\frac{1}{\sqrt{q}}\sum_{i=0}^{q-1}\ket{i}.
\end{equation}
Upon averaging over random permutations $D(x,t)$ becomes independent of the choice of the operator $\mathcal{O}_{\alpha}$, as long as it is traceless ($\tr[\mathcal{O}_{\alpha}]=0$), and conserves probability ($\bra{\flatSA}\mathcal{O}_{\alpha}=\bra{\flatSA}$). In particular for $t=0$ we have $D(x,0)=\delta_{x,0}$, while for $t>0$ the averaged decorrelator reduces to
\begin{equation} \label{eq:Dxt}
  \expval{D(x,t)}_{\rpc}=\frac{q}{q-1}-\frac{\sqrt{q}}{q-1}d(t-x+1,t+x,1),
\end{equation}
where we introduced the family of generalised diagrams $d(x,y,k)$ as (cf.\ Appendix~\ref{app:diagrammatics})
\begin{equation}
 \mkern-20mu
  d(x,y,k)=\sqrt{q}^{x+y-1}\mkern-28mu
  \begin{tikzpicture}[baseline={([yshift=-0.6ex]current bounding box.center)},scale=0.5]
    \prop{0}{0}{colPerm}{2}
    \foreach \x in {-1,1}{\prop{\x}{1}{colPerm}{2}}
    \foreach \x in {-2,0,2}{\prop{\x}{2}{colPerm}{2}}
    \foreach \x in {-3,-1,...,3}{\prop{\x}{3}{colPerm}{2}}
    \foreach \x in {-4,-2,0,2}{\prop{\x}{4}{colPerm}{2}}
    \foreach \x in {-5,-3,-1,1}{\prop{\x}{5}{colPerm}{2}}
    \foreach \x in {-6,-4,-2,0}{\prop{\x}{6}{colPerm}{2}}
    \foreach \x in {-5,-3,-1}{\prop{\x}{7}{colPerm}{2}}
    \foreach \x in {-4,-2}{\prop{\x}{8}{colPerm}{2}}
    \foreach \x in {-3}{\prop{\x}{9}{colPerm}{2}}
    \foreach \x in {0,...,6}{\circle{-0.5-\x}{-0.5+\x}}
    \foreach \x in {7,...,9}{\flatRD{-13.5+\x}{-0.5+\x}}
    \foreach \x in {1,...,3}{\circle{0.5+\x}{-0.5+\x}}
    \foreach \x in {4,...,7}{\flatLD{7.5-\x}{-0.5+\x}}
    \flatRD{0.5}{-0.5}
    \flatRD{-3.5}{9.5}
    \circle{-2.5}{9.5}
    \circle{-1.5}{8.5}
    \circle{-0.5}{7.5}
    \draw[semithick,decorate,decoration={brace}] (-6.65,6.65) -- (-3.65,9.65) node[midway,xshift=-5pt,yshift=5pt,rotate=45] {$x$};
    \draw[semithick,decorate,decoration={brace}] (0.65,6.65) -- (3.65,3.65) node[midway,xshift=5pt,yshift=5pt,rotate=-45] {$y-k$};
    \draw[semithick,decorate,decoration={brace}] (-2.35,9.65) -- (-0.35,7.65) node[midway,xshift=5pt,yshift=5pt,rotate=-45] {$k$};
  \end{tikzpicture}\mkern-5mu .
 \mkern-15mu
\end{equation}
To evaluate~\eqref{eq:Dxt} we introduce another family of diagrams
\begin{equation}
 \mkern-20mu
  z(x,y,k)=\sqrt{q}^{x+y-1}\mkern-35mu
  \begin{tikzpicture}[baseline={([yshift=0.6ex]current bounding box.center)},scale=0.5]
    \prop{0}{0}{colPerm}{2}
    \foreach \x in {-1,1}{\prop{\x}{1}{colPerm}{2}}
    \foreach \x in {-2,0,2}{\prop{\x}{2}{colPerm}{2}}
    \foreach \x in {-3,-1,...,3}{\prop{\x}{3}{colPerm}{2}}
    \foreach \x in {-4,-2,0,2}{\prop{\x}{4}{colPerm}{2}}
    \foreach \x in {-5,-3,-1,1}{\prop{\x}{5}{colPerm}{2}}
    \foreach \x in {-6,-4,-2,0}{\prop{\x}{6}{colPerm}{2}}
    \foreach \x in {-5,-3}{\prop{\x}{7}{colPerm}{2}}
    \foreach \x in {-4}{\prop{\x}{8}{colPerm}{2}}
    \foreach \x in {0,...,6}{\circle{-0.5-\x}{-0.5+\x}}
    \foreach \x in {7,...,9}{\flatRD{-13.5+\x}{-0.5+\x}}
    \foreach \x in {1,...,3}{\circle{0.5+\x}{-0.5+\x}}
    \foreach \x in {4,...,7}{\flatLD{7.5-\x}{-0.5+\x}}
    \flatRD{0.5}{-0.5}
    \circle{-3.5}{8.5}
    \circle{-2.5}{7.5}
    \circle{-1.5}{6.5}
    \circle{-0.5}{6.5}
    \draw[semithick,decorate,decoration={brace}] (-6.65,6.65) -- (-4.65,8.65) node[midway,xshift=-5pt,yshift=5pt,rotate=45] {$x-1$};
    \draw[semithick,decorate,decoration={brace}] (0.65,6.65) -- (3.65,3.65) node[midway,xshift=5pt,yshift=5pt,rotate=-45] {$y-k$};
    \draw[semithick,decorate,decoration={brace}] (-3.35,8.65) -- (-1.35,6.65) node[midway,xshift=5pt,yshift=5pt,rotate=-45] {$k$};
  \end{tikzpicture}\mkern-5mu.
  \mkern-15mu
\end{equation}
Using Eq.~\eqref{eq:WdEval2} we can show that $d(x,y,k)$, $z(x,y,k)$ fulfil exactly the same recurrence relations as $t_{\nu}(x,y,k)$, and $z_{\nu}(x,y,k)$ defined in Eqs.~\eqref{eq:relDiagramOTOCs} (cf.\ App.~\ref{sec:DiagonalOTOCs} for the precise relations and boundary conditions), when one makes the substitution $o_{\nu,2}\to {1}/{q}$.  Therefore, up to a constant factor, the averaged decorrelator has exactly the same functional form as the averaged OTOC between a diagonal and generic observable
\begin{equation}
  \expval{D(x,t)}_{\rpc}=\frac{1-{1}/{q}}{1-o_{\nu,2}}\expval{{O}_{d\nu}(x,t)}_{\rpc}.
\end{equation}

\subsection{Mutual information}
\label{sec:class2}
Here we conclude our investigation of classical dynamics generated by random permutations by computing the averaged mutual information defined Eq.~\eqref{eq:mutualinfo} after the system is prepared in a homogeneous product probability distribution in Eq.~\eqref{eq:productinitialstatemixed}.

We begin by noting that the dynamics preserves the diagonal structure of the state, and therefore also $\rho(t)$, time-evolved according to Eq.~\eqref{eq:classical-time-evolution}, is diagonal. Thus it is convenient to represent the probability distribution $\rho(t)$ as a (unnormalised) vector
\begin{equation}
\label{eq:vectorisation}
  \ket{\rho(t)}
  = q^{L} \rho(t)\ket*{\flatSA^{(1)}}^{\otimes 2L},
\end{equation}
with $\ket{\flatSA^{(1)}}$ given in Eq.~\eqref{eq:defFlatClass}. The fact that the dynamics is a permutation in the computational basis gives us two equivalent ways of viewing time-evolution: one can either evolve the state as a diagonal density matrix or as a pure state, i.e.,  
\begin{equation}
  \ket{\rho(t+1)}
  =q^{L} \rho(t+1)\ket*{\flatSA^{(1)}}^{\otimes 2L}
  =\mathbb{U}(t) 
  \ket{\rho(t)}.
\end{equation}
Using the mapping in Eq.~\eqref{eq:vectorisation}, the purity reduces to the overlap of the vector with itself
\begin{equation}
  \tr[\rho^2]= 
  \braket{\rho},
\end{equation}
while the vectorised reduced density matrices are given as
\begin{equation}
  \begin{aligned}
    \ket{\rho_A}
    &=q^{L_{\bar{A}}} (\1 \otimes \bra*{\flatSA^{(1)}}^{\otimes 2L_{\bar{A}}})
    \ket{\rho}
    ,\\
    \ket{\rho_{\bar{A}}}
    &=q^{L_{A}} (\bra*{\flatSA^{(1)}}^{\otimes 2L_{A}}\otimes \1)
    \ket{\rho},
  \end{aligned}
\end{equation}
where we assume the subsystem $A$ ($\bar{A}$) to be positioned on the left (right) and consist of $2L_{A}$ ($2 L_{\bar{A}}$) sites.

The mapping between diagonal density matrices and (unnormalised) states allows us to express the averaged subsystem purity as a $2$-replica quantity (rather than $4$-replica purity considered in Sec.~\ref{sec:entanglement}). In particular, for short times $t\leq L_A/2$ it reduces to
\begin{equation}
  \expval{\tr(\rho_A^2(t))}_{\rpc} = \gamma^{2L_A-4t} f(2t)^2,
\end{equation}
and, equivalently,
\begin{equation}
  \expval{\tr(\rho_{\bar A}^2)}_{\rpc} = \gamma^{2(L-L_A)-4t} f(2t)^2.
\end{equation}
Here we introduced $f(m)$ as
\begin{equation}
    f(m) = 
    q^{3m/2}
    \begin{tikzpicture}[baseline={([yshift=-0.6ex]current bounding box.center)},scale=0.5]
        \prop{1}{1}{colPerm}{2}
        \foreach \x in {0,2}{\prop{\x}{0}{colPerm}{2}}
        \foreach \x in {-1,1,3}{\prop{\x}{-1}{colPerm}{2}}
        \foreach \x in {-2,0,...,4}{\prop{\x}{-2}{colPerm}{2}}
        \foreach \x in {-3,...,0}{\circle{\x+0.5}{\x+1.5}}
        \foreach \x in {-3,...,0}{\flatLD{-\x+1.5}{\x+1.5}}
        \foreach \x in {-3,...,4}{\inState{\x+0.5}{-2.5}}
        \draw[semithick,decorate,decoration={brace}] (1.5,1.85) -- (4.85,-1.5) node[midway,xshift=7.5pt,yshift=7.5pt] {$m$};
    \end{tikzpicture},
\end{equation}
and introduced the symbol $\gamma$ for the overlap between the state $\ket*{\circleSA}$ and the initial state $\ket*{\inStateSA}=\ket{\rho_0}^{\otimes 2}$
\begin{equation}
  \gamma=\sqrt{q} \braket*{\circleSA}{\inStateSA}
   =\tr[\rho_0^2],\qquad
  \frac{1}{q}\le \gamma\le 1.
\end{equation}
Note that the constant $\gamma$ also parametrises the purity of the full state, as we have
\begin{equation}
  \braket{\rho(t)}=\braket{\rho(0)}=\gamma^{2L}.
\end{equation}

Using now the relations in Eq.~\eqref{eq:WdEval2}, we get the following recurrence condition for $f(m)$
\begin{equation}
    f(m)=\frac{q}{1+q} \left[\gamma^2 + \frac{1}{q}\right] f(m-1),\qquad f(0)=1,
\end{equation}
which is immediately solved by 
\be
f(m)=\left(\frac{q}{1+q} \left[\gamma^2 + \frac{1}{q}\right]\right)^m\,. 
\ee
Putting all together and recalling the definition in Eq.~\eqref{eq:mutualinfotilde} this gives 
\begin{equation}
  \tilde I(A:\bar{A})
  = -8t \log\left[\frac{q}{1+q}
  \left(\gamma+\frac{1}{q\, \gamma}\right)\right],
\end{equation}
as reported in Sec.~\ref{sec:resultsclassical}.

\section{Discussion and Outlook}
\label{sec:conclusions}
In this work we investigated what qualitative features generically associated to quantum many-body dynamics, e.g.\ growth of entanglement and operator spreading, are genuinely ``of quantum nature'', in the sense that they cannot be observed when the underlying dynamics is classical. 

To make this question well-defined and tractable we considered the specific setting of brickwork quantum circuits, where space-time is discrete and interactions are local. In this setup we compared the dynamical features of two classes of \emph{random} systems, archetypes of ``quantum'' and ``classical'' evolution. 
Specifically, we modelled generic quantum evolution using random unitary circuits~\cite{fisher2022random}, and proposed that the generic classical evolution can be faithfully represented by circuits of random, local permutations --- the classical analogue of local unitaries and a model for generic reversible microscopic dynamics. This choice is convenient because, like random unitary circuits, random permutation circuits allow for the exact analytic calculation of several probes of quantum many body dynamics. 

We compared the dynamical features of random unitary circuits to those of random permutation circuits by considering two different settings for the latter. The first is a ``quantum setting'' where everything is computed according to the rules of quantum mechanics --- although there exists a special basis where the evolution is classical and diagonal observables remain diagonal at all times. Namely, we studied quantum circuits where the local unitary gates are random permutations in the computational basis. The second setting that we considered is a true classical setting---a classical cellular automaton---where no superpositions are allowed, and only diagonal observables can be computed. In the first setting we could compare the dynamics of random unitary circuits and random permutation circuits by computing the same, standard probes, such as OTOCs and entanglement entropy, while in the second setting we considered the classical analogues of these probes.

In general we found that for generic product initial states and generic local observables the qualitative features of the dynamics generated by random permutations are completely analogous to those of random unitary circuits. More precisely, in the quantum setting we found that whenever we study the spreading of operators that are not diagonal in the ``classical'' basis or states that are generic superpositions of classical states, the phenomenology of random permutation circuits coincides with random unitary circuits. We found, however, that there are quantitative differences among the two. For example, we found that the OTOCs of generic operators in random permutation circuits (RPCs) have the same shape as those in random unitary circuits (RUCs), but they are rescaled by a constant factor smaller than one. This means in particular that their value inside the light cone is bounded below one.

Similarly, in the classical setting we found that the behaviour of decorrelators and mutual quantum information exactly mirrors that of OTOCs and entanglement entropy in random unitary circuits. In fact, we also found some exact mappings between averaged quantities computed in random permutation circuits (in both settings) and random unitary circuits. This is due to a formal equivalence between the average of two replicas of the elementary gate over random permutation and four replicas the elementary gate over random unitaries. Even though such an exact mapping is absent for a higher number of copies, we expect the behaviour of classical analogues to continue to match (at least qualitatively) quantum quantities.

Our work opens several interesting directions for future research. An immediate one is to study other aspects of the quantum many-body dynamics and the spectral properties of random permutation circuits. For instance, an obvious question that we did not consider here is to determine the classical statistical mechanical model corresponding to random permutation circuits similarly to what has been done in the random unitary case~\cite{zhou2019emergent}. Moreover, it would be interesting to understand the interplay of random permutation dynamics with measurements, investigating the occurrence of measurement induced entanglement phase transitions~\cite{skinner2019measurement,li2019measurement} and the features of the projected ensembles~\cite{cotler2023emergent} occurring in random permutation circuits. Another natural direction is to investigate the spectral properties of Floquet random permutation circuits. Indeed, it is interesting to wonder whether the latter will be described by a classic random matrix ensemble as it is the case for random unitary circuits~\cite{chan2018spectral,friedman2019spectral, garratt2021manybody, garratt2021local}. Finally, our techniques might also find application in problems related to encryption~\cite{chamon2022quantum}.

A more general question concerns the quantum computational capabilities of random permutation circuits. Indeed, random unitary circuits are known to form approximate unitary $k$ designs~\cite{harrow2009random, brandao2016efficient, brandao2016local, hunterjones2019unitary, haferkamp2022randomquantum, haferkamp2023efficient}, i.e.\ polynomial-depth circuits of random unitaries reproduce the moments of a many-body random matrix with arbitrary precision. Analogously, random permutation circuits are known to form approximate permutation $k$ designs~\cite{gowers1996an, hoory2005simple, hoory2005simple, he2024pseudorandom, chen2024incompressibility}, i.e.\ polynomial-depth circuits of random permutations reproduce the moments of a many-body permutation matrix with arbitrary precision. Our results on the behaviour of quantum information in these circuits, however, suggests that hiding the information about the privileged basis, e.g.\ via a small number of random local basis changes, might be sufficient to lift random permutations to approximate unitary $k$ designs. Interestingly, a concrete construction to realise unitary $k$ designs using random permutations has been recently presented in Ref.~\cite{chen2024efficientunitarydesignspseudorandom}.

\begin{acknowledgments}
  We thank Toma\v{z} Prosen and Ignacio Cirac for useful discussions. We acknowledge financial support from the Royal Society through the University Research Fellowship No.\ 201101 (B.\ B.), the Leverhulme Trust through the Early Career Fellowship No.\ ECF-2022-324 (K.\ K.), Alexander von Humboldt Foundation (P.\ K.), and the Novo Nordisk Foundation under grant numbers NNF22OC0071934 and NNF20OC0059939 (D.\ M.). B.\ B., K.\ K., and P.\ K.\ warmly acknowledge the hospitality of the University of Ljubljana where this work has been finalised. 
\end{acknowledgments}

\onecolumngrid

\appendix

\section{Summary of diagrammatic notation}
\label{app:diagrammatics}

Here we summarise the diagrammatic notation used throughout the manuscript. As explained in Sec.~\ref{sec:setting}, we work in a \emph{folded} picture, where each line represents $n$ copies of the one-qudit Hilbert space, and the two-site gates are represented by big squares (cf.\ Eq.~\ref{eq:defReplicas})
\begin{equation}
  \begin{tikzpicture}[baseline={([yshift=-0.6ex]current bounding box.center)},scale=.75]
    \prop{0}{0}{colU}{2n}
  \end{tikzpicture}
  = \left(U(x,t)\otimes U(x,t)^{\ast}\right)^{\otimes n}.
\end{equation}
Here $U(x,t)$ is time and space dependent, which is represented by using different shades of blue, and no label at the centre of the square implies a single copy. 
In this representation, connected legs imply summation over all the basis states, and the matrix multiplication is understood to go from bottom to top.

We represent averaged gates with different colours (cf.\ Eqs.\ \eqref{eq:explicitaverageruc}, \eqref{eq:permave}, and~\eqref{eq:P}), 
\begin{equation}
  \begin{tikzpicture}[baseline={([yshift=-0.6ex]current bounding box.center)},scale=.75]
    \prop{0}{0}{colIHaar}{2n}
  \end{tikzpicture}=\mathcal{U}^{(2n)}
  =\expval*{\left(U\otimes U^{\ast}\right)^{\otimes n}}_{\ruc},\qquad
  \begin{tikzpicture}[baseline={([yshift=-0.6ex]current bounding box.center)},scale=.75]
    \prop{0}{0}{colPerm}{n}
  \end{tikzpicture}=\mathcal{Q}^{(n)}=\expval*{U^{\otimes n}}_{\rpc},\qquad
  \begin{tikzpicture}[baseline={([yshift=-0.6ex]current bounding box.center)},scale=.75]
    \prop{0}{0}{white}{2n}
  \end{tikzpicture}=\expval*{\left(U\otimes U^{\ast}\right)^{\otimes n}},
\end{equation}
where the transparent colour (right-most diagram) represents an averaged gate without specifying the group over which it is averaged (i.e.\ could be either RUC or RPC).

Various shapes with one leg sticking out represent vectors in the (replicated) Hilbert space, with the superscript denoting the number of replicas. In particular, empty circle, square, cross, and bar represent different (normalised) partition states,
\begin{equation}
  \begin{aligned}
  \begin{tikzpicture}[baseline={([yshift=-0.6ex]current bounding box.center)},scale=0.75]
    \gridLine{-0.3}{-0.3}{-0.3}{0.45}
    \circle{-0.3}{-0.3}
    \node[scale=0.7] at (-0.05,-.35) {$2$};
  \end{tikzpicture}&
    =\ket*{\circleSA^{(2)}}
    =\frac{1}{\sqrt{q}} \ket{\{\{0,1\}\}},&\qquad
  \begin{tikzpicture}[baseline={([yshift=-0.6ex]current bounding box.center)},scale=0.75]
    \gridLine{-0.3}{-0.3}{-0.3}{0.45}
    \circle{-0.3}{-0.3}
    \node[scale=0.7] at (-0.05,-.35) {$4$};
  \end{tikzpicture}&
    =\ket*{\circleSA^{(4)}}
    =\frac{1}{q} \ket{\{\{0,1\},\{2,3\}\}},\\
  \begin{tikzpicture}[baseline={([yshift=-0.6ex]current bounding box.center)},scale=0.75]
    \gridLine{-0.3}{-0.3}{-0.3}{0.45}
    \flat{-0.3}{-0.3}
    \node[scale=0.7] at (-0.05,-.35) {$2$};
  \end{tikzpicture}&
    =\ket*{\flatSA^{(2)}}
    =\frac{1}{q} \ket{\{\{0\},\{1\}\}},&\qquad
  \begin{tikzpicture}[baseline={([yshift=-0.6ex]current bounding box.center)},scale=0.75]
    \gridLine{-0.3}{-0.3}{-0.3}{0.45}
    \flat{-0.3}{-0.3}
    \node[scale=0.7] at (-0.05,-.35) {$4$};
  \end{tikzpicture}&
    =\ket*{\flatSA^{(4)}}
    =\frac{1}{q^2} \ket{\{\{0\},\{1\},\{2\},\{3\}\}},\\
  \begin{tikzpicture}[baseline={([yshift=-0.6ex]current bounding box.center)},scale=0.75]
    \gridLine{-0.3}{-0.3}{-0.3}{0.45}
    \square{-0.3}{-0.3}
    \node[scale=0.7] at (-0.05,-.35) {$4$};
  \end{tikzpicture}&
    =\ket*{\squareSA^{(4)}}
    =\frac{1}{q} \ket{\{\{0,3\},\{1,2\}\}}, &\qquad 
  \begin{tikzpicture}[baseline={([yshift=-0.6ex]current bounding box.center)},scale=0.75]
    \gridLine{-0.3}{-0.3}{-0.3}{0.45}
    \comb{-0.3}{-0.3}
    \node[scale=0.7] at (-0.05,-.35) {$4$};
  \end{tikzpicture}&
    =\ket*{\combSA^{(4)}}
    =\frac{1}{\sqrt{q}} \ket{\{\{0,1,2,3\}},
  \end{aligned}
\end{equation}
see App.~\ref{app:partitions} for a precise definition. Note that $\ket*{\circleSA^{(2)}}$, $\ket*{\circleSA^{(4)}}$, and $\ket*{\squareSA^{4}}$ can be equivalently understood as permutation states $\ket{\sigma}$ defined in Eq.~\eqref{eq:coefficientsperm}, which are invariant also under random unitaries. In particular, for $2n=2$ there is only one invariant state corresponding to the only permutation $\sigma_1$ of $1$ element,
\begin{equation}
  \sigma_1(1)=1,\qquad
  \ket{\sigma_1}=\sum_{a=0}^{q-1} \ket{a a} = \sqrt{q} \ket*{\circleSA^{(2)}}. 
\end{equation}
In the $2n=4$ case, there are two permutations, $\{\sigma_1,\sigma_2\}$, corresponding to identity and transposition,
\begin{equation}
  \sigma_1(1)=1,\ \sigma_1(2)=2,\qquad \sigma_2(1)=2,\ \sigma_2(2)=1,
\end{equation}
which give the states that (up to normalisation) coincide with $\ket*{\circleSA^{(4)}}$, and $\ket{\squareSA^{(4)}}$,
\begin{equation}
  \ket{\sigma_1}=\sum_{a=0}^{q-1}\sum_{b=0}^{q-1} \ket{aabb}= q\ket*{\circleSA^{(4)}},\qquad
  \ket{\sigma_2}=\sum_{a=0}^{q-1}\sum_{b=0}^{q-1} \ket{abba}= q\ket*{\squareSA^{(4)}}.
\end{equation}
It is straightforward to see that these two states satisfy the defining condition~\eqref{eq:coefficientsperm}. Note that throughout the manuscript we often suppress the subscripts (and superscripts) denoting the number of replica, if the latter is clear from the context.

Solid circles and squares are used to denote a one-site operator $\mathcal{O}_{\mu}$ applied to the relevant partition states (see Eqs.~\eqref{eq:blackcirclestate}, and \eqref{eq:blackcirclesquare4}),
\begin{equation}
  \begin{tikzpicture}[baseline={([yshift=-0.6ex]current bounding box.center)},scale=0.75]
    \gridLine{-0.3}{-0.3}{-0.3}{0.45}
    \blackcircle{-0.3}{-0.3}
    \node[scale=0.7] at (-0.05,-.35) {$2$};
    \node at (-0.05,-0.75) {\scalebox{0.9}{$\mathcal{O}_{\mu}$}};
  \end{tikzpicture}=\mathcal{O}_{\mu}\otimes \1
  \ket*{\circleSA^{(2)}},\qquad
  \begin{tikzpicture}[baseline={([yshift=-0.6ex]current bounding box.center)},scale=0.75]
    \gridLine{-0.3}{-0.3}{-0.3}{0.45}
    \blackcircle{-0.3}{-0.3}
    \node[scale=0.7] at (-0.05,-.35) {$4$};
    \node at (-0.05,-0.75) {\scalebox{0.9}{$\mathcal{O}_{\mu}$}};
  \end{tikzpicture}=\left(\mathcal{O}_{\mu}\otimes \1\right)^{\otimes 2}
  \ket*{\circleSA^{(4)}},\qquad
  \begin{tikzpicture}[baseline={([yshift=-0.6ex]current bounding box.center)},scale=0.75]
    \gridLine{-0.3}{-0.3}{-0.3}{0.45}
    \blacksquare{-0.3}{-0.3}
    \node[scale=0.7] at (-0.05,-.35) {$4$};
    \node at (-0.05,-0.75) {\scalebox{0.9}{$\mathcal{O}_{\mu}$}};
  \end{tikzpicture}=\left(\mathcal{O}_{\mu}\otimes \1\right)^{\otimes 2}
  \ket*{\squareSA^{(4)}},
\end{equation}
where we once again often suppress the superscript $2/4$ if the number of
replicas is clear from the context.

Finally, we use solid triangles to denote various initial states. In particular, in Fig.~\ref{fig:statet} the triangles represent the (possibly disordered) initial state,
\begin{equation}
  \begin{tikzpicture}[baseline={([yshift=-0.6ex]current bounding box.center)},scale=0.75]
    \gridLine{-0.3}{-0.3}{-0.3}{0.45}
    \inState{-0.3}{-0.3}
  \end{tikzpicture}=\ket{\psi_j},
\end{equation}
while in Sec.~\ref{sec:entanglement} we use it for the average over four copies of the initial state,
\begin{equation}
  \begin{tikzpicture}[baseline={([yshift=-0.6ex]current bounding box.center)},scale=0.75]
    \gridLine{-0.3}{-0.3}{-0.3}{0.45}
    \inState{-0.3}{-0.3}
  \end{tikzpicture}=q\expval*{\left(\ket{\psi}\otimes\ket{\psi}^{\ast}\right)^{\otimes 2}}.
\end{equation}
Similarly, in Sec.~\ref{sec:class2} the same symbol is used to denote the two copies of the vectorized initial density matrix (cf.\ Eqs.\ \eqref{eq:productinitialstatemixed}, and \eqref{eq:vectorisation}),
\begin{equation}
  \begin{tikzpicture}[baseline={([yshift=-0.6ex]current bounding box.center)},scale=0.75]
    \gridLine{-0.3}{-0.3}{-0.3}{0.45}
    \inState{-0.3}{-0.3}
  \end{tikzpicture}=\ket{\rho_0}^{\otimes 2}.
\end{equation}

\section{Partitions for $n=1,2,3,4$}
\label{app:partitions}

In this appendix we list all states $\{\ket{\pi_{j}}\}_{j=1}^{B_n}$ for $n=1,2,3,4$. For $n=1$ we have only one partition, and the corresponding vector is
\begin{equation}
  \ket{\pi_1}=\ket{\{\{0\}\}}=\sum_{a=0}^{q-1} \ket{a}.
\end{equation}
For $n=2$ there are two distinct partitions corresponding to the vectors
\begin{equation}
  \ket{\pi_{1}}=\ket{\{\{0,1\}\}}=\sum_{a=0}^{q-1} \ket{aa},\qquad
  \ket{\pi_{2}}=\ket{\{\{0\},\{1\}\}}=\sum_{a,b=0}^{q-1} \ket{ab}.
\end{equation}
For $n=3$ we have $5$ distinct partitions corresponding to the vectors 
\begin{equation}
  \begin{aligned}
    \ket{\pi_{1}}&=\ket{\{\{0,1,2\}\}}=\sum_{a=0}^{q-1}\ket{aaa},
    &\quad
    \ket{\pi_{2}}&=\ket{\{\{0,1\},\{2\}\}}=\sum_{a,b=0}^{q-1}\ket{aab},
    \\
    \ket{\pi_{3}}&=\ket{\{\{0,2\},\{1\}\}}=\sum_{a,b=0}^{q-1}\ket{aba},
    &\quad
    \ket{\pi_{4}}&=\ket{\{\{1,2\},\{0\}\}}=\sum_{a,b=0}^{q-1}\ket{abb},
    \\
    \ket{\pi_{5}}&=\ket{\{\{0\},\{1\},\{2\}\}}=\sum_{a,b,c=0}^{q-1}
    \ket{abc},
  \end{aligned}
\end{equation}
For $n=4$, there are $15$ distinct partitions corresponding to the vectors
\begin{equation}
  \begin{aligned}
    \ket{\pi_{1}}&=\ket{\{\{0,1,2,3\}\}}=\sum_{a=0}^{q-1}\ket{aaaa},
    &\quad
    \ket{\pi_{2}}&=\ket{\{\{0,1,2\},\{3\}\}}=\sum_{a,b=0}^{q-1}
    \ket{aaab},
    \\
    \ket{\pi_{3}}&=\ket{\{\{0,1,3\},\{2\}\}}=\sum_{a,b=0}^{q-1}
    \ket{aaba},
    &\quad
    \ket{\pi_{4}}&=\ket{\{\{0,2,3\},\{1\}\}}=\sum_{a,b=0}^{q-1}
    \ket{abaa},
    \\
    \ket{\pi_{5}}&=\ket{\{\{1,2,3\},\{0\}\}}=\sum_{a,b=0}^{q-1}
    \ket{abbb},
    &\quad
    \ket{\pi_{6}}&=\ket{\{\{0,1\},\{2,3\}\}}=\sum_{a,b=0}^{q-1}
    \ket{aabb},
    \\
    \ket{\pi_{7}}&=\ket{\{\{0,3\},\{1,2\}\}}=\sum_{a,b=0}^{q-1}
    \ket{abba},
    &\quad
    \ket{\pi_{8}}&=\ket{\{\{0,2\},\{1,3\}\}}=\sum_{a,b=0}^{q-1}
    \ket{abab},
    \\
    \ket{\pi_{9}}&=\ket{\{\{0,1\},\{2\},\{3\}\}}=
    \sum_{a,b,c=0}^{q-1} \ket{aabc},
    &\quad
    \ket{\pi_{10}}&=\ket{\{\{0,2\},\{1\},\{3\}\}}=
    \sum_{a,b,c=0}^{q-1} \ket{abac},
    \\
    \ket{\pi_{11}}&=\ket{\{\{0,3\},\{1\},\{2\}\}}=
    \sum_{a,b,c=0}^{q-1} \ket{abca},
    &\quad
    \ket{\pi_{12}}&=\ket{\{\{1,3\},\{0\},\{2\}\}}=
    \sum_{a,b,c=0}^{q-1} \ket{abcb},
    \\
    \ket{\pi_{13}}&=\ket{\{\{1,2\},\{0\},\{3\}\}}=
    \sum_{a,b,c=0}^{q-1} \ket{abbc},
    &\quad
    \ket{\pi_{14}}&=\ket{\{\{2,3\},\{0\},\{1\}\}}=
    \sum_{a,b,c=0}^{q-1} \ket{abcc},
    \\
  \ket{\pi_{15}}&=\ket{\{\{0\},\{1\},\{2\},\{3\}\}}=
  \sum_{a,b,c,d=0}^{q-1} \ket{abcd}.
  \end{aligned}
\end{equation}

\section{Solution to the recurrence relation}
\label{app:recSol}
Here we derive the solution to the recurrence equations defined in Eq.~\eqref{eq:recOriginal}. Our method is a variant of the so called kernel method~\cite{pemantle2013analytic} and is applicable for any boundary condition. 

We begin by simplifying the system of equations. Specifically, we express $v_{k}(x,y)$ in terms of $w_{k}(x,y)$ using the first of Eq.~\eqref{eq:recOriginal} and introduce 
\begin{equation}
  u_k(x,y)=\left(\frac{1+q}{\sqrt{q}}\right)^{x+2k+2y}
  \left[w_{x}(x+k,y)-q^{-\frac{x+1}{2}}\right].
  \label{eq:changeofvariableu}
\end{equation}
This allows us to write Eq.~\eqref{eq:recOriginal} as a single recurrence relation
\begin{equation}
\label{eq:singleRecRelation}
  u_{k}(x,y)
  =u_{k}(x-1,y)
  +\left(1+\frac{1}{q}\right)u_0(x+k,y-1) +\sum_{r=1}^{k}u_{r}(x+k-r,y-1)
\end{equation}
with boundary conditions given by
\begin{equation}
\label{eq:singleRecRelationBCs}
  u_{k}(0,y)=0,\qquad
  u_k(x,0)=\delta_{k,0}(1-\delta_{x,0})
  \frac{q-1}{\sqrt{q}}\left(1+\frac{1}{q}\right)^x.
\end{equation}
From now on, however, we will take the boundary conditions to be general and set 
\begin{equation}
\label{eq:genBC}
  u_{k}(0,y)=\psi_{k,y},\qquad
  u_{k}(x,0)=\phi_{k,x},\qquad
  \phi_{k,0}=\psi_{k,0}.
\end{equation}
Note that in these new variables the function of interest reads as 
\begin{equation}
  w_{1}(x,y)= \frac{1}{q}+\left(\frac{\sqrt{q}}{1+q}\right)^{2x+2y-1}
  u_{x-1}(1,y).
\end{equation}

Equation~\eqref{eq:singleRecRelation} can be simplified further to a form involving only differences of neighbouring terms. To achieve this we subtract the equation evaluated at $x+1$ and $k-1$ from the one evaluated at $x$ and $k$. This gives
\begin{equation}
\label{eq:IMRecRel}
  u_{k}(x,y)  +u_{k-1}(x,y)  -u_{k}(x,y-1)  -u_{k}(x-1,y) -u_{k-1}(x+1,y) =0,
\end{equation}
which holds for all $k,x,y\ge 1$, while for $k=0$ we have
\begin{equation}
\label{eq:IMRecRel2}
  u_{0}(x,y)=u_{0}(x-1,y)+\alpha u_{0}(x,y-1),\qquad
  \alpha:=1+\frac{1}{q}. 
\end{equation}

Next, we proceed to solve \eqref{eq:IMRecRel}, \eqref{eq:IMRecRel2}, with the boundary conditions \eqref{eq:genBC}. First we observe that \eqref{eq:IMRecRel2} is a single difference equation for $u_{0}(x,y)$ and can be explicitly solved by iteration. The solution reads as
\begin{equation}
  u_{0}(x,y)=
  \alpha^y \sum_{m=0}^{x-1}\binom{y+m-1}{m}\phi_{0,x-m}
  +\sum_{n=0}^{y-1}\alpha^n\binom{x+n-1}{n}\psi_{0,y-n}
  +\delta_{x,0}\delta_{y,0}\phi_{0,0}.
\end{equation}
Then, to solve \eqref{eq:IMRecRel}, we define a rescaled \emph{generating function} $g_{k,y}(z)$ as
\begin{equation}
    g_{k,y}(z)=z^k(1-z)^y \sum_{x=1}^{\infty} z^x u_{k}(x,y).
\end{equation}
Eq.~\eqref{eq:IMRecRel} can then be translated in the following equation for the generating function 
\begin{equation}\label{eq:recurrenceGtilde}
   {g}_{k,y}-{g}_{k-1,y}-{g}_{k,y-1}
  -z^{k+1}(1-z)^{y-1} \psi_{k,y}
  +z^k(1-z)^{y-1}u_{k-1}(1,y)
  =0
\end{equation}
with the boundary conditions
\begin{equation}\label{eq:recurrenceGtildeBC}
  {g}_{k,0}(z)=\sum_{x=1}^{\infty} z^{x+k}\phi_{k,x},\qquad
  {g}_{0,y}(z)=
    \alpha^y
    \sum_{m=1}^{\infty}
    z^m \phi_{0,m}
  + \sum_{n=1}^{y}
    z(1-z)^{n-1}\alpha^{y-n}\psi_{0,n}.
\end{equation}
In this way we have reduced the recurrence relation in three
variables into a recurrence relation in only two variables (with an explicit dependence on the parameter $z$) and driving term
\be
z^{k+1}(1-z)^{y-1} \psi_{k,y} - z^k(1-z)^{y-1}u_{k-1}(1,y).
\ee
This relation is a linear difference equation, therefore its
solution can be written as 
\begin{equation}
  {g}_{k,y}(z)={g}^{(1)}_{k,y}(z)+{g}^{(2)}_{k,y}(z),
\end{equation}
where ${g}^{(1)}_{k,y}$ solves~\eqref{eq:recurrenceGtilde} with simpler
boundary conditions ${g}^{(1)}_{k,0}(z)={g}^{(1)}_{0,y}(z)=0$. The latter can again be solved by iteration to give  
\begin{equation}
  {g}^{(1)}_{k,y}(z)=
  -\sum_{m=1}^{k}\sum_{n=1}^{y}
  \binom{k-m+y-n}{y-n}
  z^m(1-z)^{n-1}
  \left(u_{m-1}(1,n)- z\psi_{m,n}\right).
\end{equation}
On the other hand, ${g}^{(2)}_{k,y}(z)$ is a solution to the homogeneous equation with the boundary conditions given by~\eqref{eq:recurrenceGtildeBC}. Solving by iteration once again we find 
\begin{equation}
  {g}^{(2)}_{k,y}(z)=
  \sum_{m=1}^{k}\binom{y-1+k-m}{y-1}{g}_{m,0}(z)
  +\sum_{n=1}^{y}\binom{k-1+y-n}{k-1}{g}_{0,n}(z)
  +\delta_{k,0}\delta_{y,0}{g}_{0,0}(z).
\end{equation}
Here we have used the values of $u_x(1,y)$ as independent parameters. However, they are not arbitrary, as by construction the function $g_{k,y}(z)$ has to be analytic in the neighbourhood of $z=0$ and its derivatives give the values of $u_{k}(x,y)$
\begin{equation}
\label{eq:conditionsgen}
  g_{k,y}(0)=0,\qquad \frac{{\rm d}^m}{{{\rm d}z}^m} \frac{g_{k,y}(z)}{z^k(1-z)^y}\biggl|_{z=0} = m!\cdot u_{k}(m,y). 
\end{equation}
This gives us a set of conditions on $u_x(1,y)$, which can be used to fully determine it. This idea is the key step of the kernel method~\cite{pemantle2013analytic}. Whenever $u_x(1,y)$ are finite, which can be verified a posteriori, this method leads to a consistent solution.

To find a close form expression we note that  Eqs.~\eqref{eq:conditionsgen} immediately imply that the first $k$ derivatives of ${g}_{k,y}$ in $z=0$ should vanish, i.e.\ 
\begin{equation}
 \frac{{\rm d}^l}{{{\rm d}z}^l} {g}_{k,y}(0)=0,\qquad\qquad 0\leq l\leq k. 
\end{equation}
This gives us a set of $k$ conditions for $\{u_n(1,m)\}_{n=0,\ldots,k-1; 1,\ldots,y}$ for each $y$ and $k$. We write them as follows
\begin{equation}
\label{eq:abcconditions}
  a(l,k,y)=b(l,k,y)+c(l,k,y),\qquad\qquad 0\leq l\leq k,
\end{equation}
where we set 
\begin{equation}
  \begin{aligned}
    a(l,k,y)&=\sum_{m=1}^{k}\sum_{n=1}^{y}
    \binom{k-m+y-n}{y-n}
    (-1)^{l-m} 
    \binom{n-1}{l-m} u_{m-1}(1,n),\\
    b(l,k,y)&= \sum_{m=1}^{k}\sum_{n=1}^{y} \binom{k-m+y-n}{y-n}
    (-1)^{l-m-1} \binom{n-1}{l-m-1}\psi_{m,n}\\
    &+\sum_{n=1}^{y} \binom{n-1}{l-1} (-1)^{l-1} \psi_{0,n}
    \left(A_{k,y-n}(\alpha)+\binom{k-1+y-n}{k-1}\right),\\
    c(l,k,y)&=\sum_{m=1}^{l-1}\binom{y-1+k-m}{y-1} \phi_{m,l-m} + \phi_{0,l}  A_{k,y}(\alpha),
  \end{aligned}
\end{equation}
and introduced the auxiliary function
\begin{equation}\label{eq:defAky2}
  A_{k,y}(z)=\sum_{t=1}^{y}\binom{k-1+y-t}{k-1} z^t +\delta_{y,0}\delta_{k,0}.
\end{equation}
Equations~\eqref{eq:abcconditions} are enough to determine $\{u_n(1,m)\}_{n=0,\ldots,k-1; 1,\ldots,y}$. To show it we explicitly invert them using the following binomial identity (proven in App.~\ref{sec:proofinvid})
\begin{lemma}
\label{lemma:binomialindentities}
For $m, y\ge1$ and $1\le n\le y$ 
\be
\sum_{\ell =m}^{k}
  (-1)^{\ell -m} \binom{n-1}{\ell -m} \binom{y-2+k-\ell}{y-2}=\binom{y-n-1+k-m}{y-n-1}. 
\label{eq:binomialidentity}
\ee
\end{lemma}

\noindent
Specifically, by repeated use of \eqref{eq:binomialidentity} we find  
\begin{equation}
\label{eq:rightLinearCombinationU1}
    \sum_{l=1}^k \left(\binom{y-2+k-l}{y-2} a(l,k,y)
   -\binom{y-1+k-l}{y-1}a(l,k,y-1)\right)
    = u_{k-1}(1,y),
\end{equation}
which gives us the explicit form of $u_{k-1}(1,y)$
\begin{equation}
  u_{k-1}(1,y)=
  \sum_{l=1}^k \left(\binom{y-2+k-l}{y-2} [b(l,k,y)+c(l,k,y)]
   -\binom{y-1+k-l}{y-1}[b(l,k,y-1)+c(l,k,y-1)]\right).
\end{equation}
The r.h.s.\ can be again simplified using \eqref{eq:binomialidentity}. The final form of the solution for generic boundary conditions reads as 
\begin{equation}
  \begin{aligned}
   u_{k-1}&(1,y) = \delta_{k,1}(1-\delta_{y,0})\psi_{0,y}
    +\sum_{n=1}^{y-1}\psi_{0,n}
    \left(\binom{y-n-2+k}{k-1} A_{k,y-n}(\alpha)
    - \binom{y-n-1+k}{k-1} A_{k,y-1-n}(\alpha)\right)\\
    &+
    \sum_{m=1}^{k-1}
    \sum_{n=1}^y
    \left[
      \binom{k-m+y-n-2}{y-n-1}\binom{k-m+y-n}{y-n}-
      \binom{k-m+y-n-1}{y-n}\binom{k-m+y-n-1}{y-n-1}
    \right] \psi_{m,n}\\
    & + \sum_{l=1}^k \sum_{m=1}^{l}
    \left[ \binom{y-2+k-l}{y-2} \binom{y-1+k-m}{y-1} -
    \binom{y-1+k-l}{y-1} \binom{y-2+k-m}{y-2} \right]
    \phi_{m,l-m} \\
    &+A_{k,y}(\alpha)B_{y-1,k}-A_{k,y-1}(\alpha)B_{y,k},
  \end{aligned}
\end{equation}
where we introduced the auxiliary
function 
\begin{equation}
  B_{k,y}=\sum_{t=1}^y\binom{k-1+y-t}{k-1}\phi_{0,t}+\delta_{y,0}\delta_{k,0}\phi_{0,0}.
\end{equation}
Specialising it to the boundary conditions \eqref{eq:singleRecRelationBCs} we find
\begin{equation}\label{eq:solution2pcorr}
  u_{k-1}(1,y)=\frac{q-1}{\sqrt{q}} \left(A_{k,y}(\alpha)A_{y-1,k}(\alpha)-A_{k,y-1}(\alpha)A_{y,k}(\alpha)\right).
\end{equation}
Plugging into Eq.~\eqref{eq:changeofvariableu} we find Eq.~\eqref{eq:w1solution}.

\subsection{Proof of Lemma~\ref{lemma:binomialindentities}}
\label{sec:proofinvid}

To prove Eq.~\eqref{eq:binomialidentity} we begin by rewriting the l.h.s.\ as follows 
\begin{equation}
  \sum_{\ell =m}^{k}
  (-1)^{\ell -m} \binom{n-1}{\ell -m} \binom{y-2+k-\ell}{y-2}=
  \sum_{r=0}^{n-1}
  (-1)^r \binom{n-1}{r} \binom{y-2+k-m-r}{y-2},
\end{equation}
which follows from from a change of variables from $\ell$ to $r=\ell-m$ and the observation 
\begin{equation}
  \left.\binom{n-1}{r}\right|_{r\ge n}=0,\qquad 
  \left.\binom{y-2+k-m-r}{y-2}\right|_{r \ge k-m}=0,\qquad y\geq 1\,.
\end{equation}
Introducing now
\begin{equation}
  f_{r}(k,y,m)=\binom{y-2+k-m-r}{y-2},
\end{equation}
we can interpret the above expression as a binomial transform
of the sequence $\{f_r(k,y,m)\}_{r}$~\cite{bernstein1995some}, which is equivalent to the $n-1$-th forward difference of the sequence computed in $r=0$. More precisely we have 
\begin{equation}
  \sum_{r=0}^{n-1}(-1)^r\binom{n-1}{r}  f_{r}(k,y,m) = (-1)^{n-1}
  \left(\Delta^{n-1}f(k,y,m)\right)_{0},\qquad
  (\Delta h)_{j}:=h_{j+1}-h_j.
\end{equation}
Using now the binomial recursive relation for $f_r(k,y,m)$ we get
\begin{equation}
  (\Delta f(k,y,m))_r=-f(k,y-1,m)_r,
\end{equation}
which in particular implies that the $(n-1)$-th difference gives
\begin{equation}
  (\Delta^{n-1} f(k,y,m))_r=(-1)^{n-1} f_r(k,y-n+1,m).
\end{equation}
Putting all together we find   
\be
  \sum_{\ell =m}^{k}
  (-1)^{\ell -m} \binom{n-1}{\ell -m} \binom{y-2+k-\ell}{y-2} = \binom{y-n-1+k-m}{y-n-1}. 
\ee
\qed

\section{Asymptotic form of $A_{x,y}(z)$}
\label{sec:asyA}

Here we give the asymptotic scaling of
\begin{equation}
  A_{x,y}(z)=\sum_{r=1}^{y}\binom{x-1+y-r}{x-1} z^r +\delta_{y,0}\delta_{x,0}.
\end{equation}
Approximating the binomial symbol using the Stirling's approximation, and evaluating the sum via saddle-point, we get the following expression valid at leading order in $l=x+y$ for fixed $\eta=x/(x+y)$
\begin{equation}
  \left. A_{x,y}(z)\right|_{x=l \eta, y=l(1-\eta)} \simeq \begin{cases}
    z^{l}(z-1)^{-l\eta}
  ,
    &\eta<\frac{z-1}{z},\\
    (1-\eta)^{\frac{1}{2}-l(1-\eta)} \eta^{\frac{1}{2}-l\eta}
    \frac{z}{\sqrt{8 \pi l}}
    \left(1+\frac{2}{\log\frac{1}{z(1-\eta)}}\right),\ &\eta>\frac{z-1}{z},
\end{cases}
\label{eq:Aasy}
\end{equation}
while close to the transition point we have
\begin{equation}
  \left.\frac{(z-1)^x}{z^{x+y}}A_{x,y}(z)
  \right|_{x\approx \frac{z-1}{z}(x+y)+\xi \sqrt{x+y}} \simeq
  \frac{1}{2}\left[1-\mathrm{erf}\frac{z\xi}{\sqrt{2(z-1)}}\right].
\end{equation}

\subsection{Correlation functions}
Using the above expressions one can directly evaluate the asymptotic expansion of dynamical correlations via Eq.~\eqref{eq:avecorrresult}. In this case the function $A_{x,y}(z)$ is evaluated in $z=\alpha=1+{1}/{q}$ while the averaged correlation function reads as
\be
\label{eq:Correff}
 \expval{C_{\mu\nu}(x,t)}_{\rpc} = o_{\mu} o_{\nu} f(t-x+1,t+x),
\ee
where we introduced 
\begin{equation}
  f(x,y)=\frac{(\alpha-1)^{x+y-1}}{\alpha^{2(x+y)-1}(2-\alpha)}
  \left(A_{x,y}(\alpha)A_{y-1,x}(\alpha)-A_{x,y-1}(\alpha)A_{y,x}(\alpha)\right).
\end{equation}
To understand what cases of Eq.~\eqref{eq:Aasy} to consider we then have to compare 
\be
\frac{x}{x+y}=\eta\in[0,1], 
\ee
with $1-1/z$. Since $1-1/z\in[0,1/2]$ we have three cases: (i) $\eta\leq 1-1/z$, (ii) $1-1/z\leq \eta\leq 1/z$, and (iii) $\eta\leq 1/z$. The leading contribution is attained in the second case, when both $\eta$ and $1-\eta$ are larger than $1-1/z$ and $f(x,y)$ can be approximated by an exponentially decaying Gaussian centred around $x/(x+y)=y/(x+y)=1/2$
\begin{equation}
  \left.f(\eta l,(1-\eta)l)\right|_{\eta,1-\eta>(\alpha-1)/\alpha}\simeq
  \left(\frac{4(\alpha-1)}{\alpha^2}\right)^{l-1}
  \exp[- l (1-2\eta)^2]
  \frac{2+\log\frac{2}{\alpha}}{4\pi l^2(\frac{2}{\alpha}-1) \log^3\frac{2}{\alpha}}.
\end{equation}
This expression does not account for the tails, which should be evaluated in the other regimes. Plugging back into \eqref{eq:Correff} and retaining only leading order terms this recovers Eq.~\eqref{eq:correlationasy} with  
\begin{equation} \label{eq:Kmunu}
  K_{\mu\nu} = o_{\mu}o_{\nu}
  \frac{(2+\log\frac{2q}{1+q})}{16\pi(\frac{q-1}{q+1}) \log^3\frac{2q}{1+q}}\,.
\end{equation}

\subsection{OTOCs}
To find the asymptotic expansion of averaged OTOCs we recall  
\be
\expval{{O}_{\mathrm{d},\gamma}(x,t)}_{\rpc}=
  (1-o_{\gamma,2}) g(t-x+1,t+x)
  \label{eq:OTOCeff}
\ee
with 
\begin{equation}
  g(x,y)=
  \frac{(z-1)^{x+y}}{(z-2)z^{2(x+y)-1}}
  \left(A_{x,y}(z)A_{y-1,x}(z)-A_{x,y-1}(z)A_{y,x}(z)\right),
\end{equation}
and that $z= \beta = 1+q\ge 3$. 

Also in this case we have three possible regimes for $x/(x+y)=\eta$: (i) $\eta\leq 1/z$, (ii) $1/z\leq \eta\leq 1-1/z$, and (iii) $\eta\leq 1-1/z$. The only contribution that is not exponentially suppressed is found in the second regime and is given by 
\begin{equation}
  \left.g(\eta l,(1-\eta)l)\right|_{\eta,1-\eta<1-1/z+\mathcal{O}(\frac{1}{l})}
  \simeq 
 \frac{1}{4} \left({1+\erf\left[\frac{z-z\eta-1}{\sqrt{2(z-1)/l}}\right]}\right)
  \left({1+\erf\left[\frac{z\eta-1}{\sqrt{2(z-1)/l}}\right]}\right).
\end{equation}
Plugging back into \eqref{eq:OTOCeff} and retaining only leading order contributions this expression gives Eq.~\eqref{eq:OTOCasy}.

\section{Averaged OTOCs in Random Permutation Circuits}
\label{sec:DiagonalOTOCs}
Here we consider an OTOC between two observables, of which (at least) one is diagonal. In this case we can use the fact that in the case of a traceless diagonal Hermitian operator $\mathcal{O}_{\rm d}$, the projection of $\ket*{\blackcircleSA{\mathrm{d}}^{(4)}}$ (cf.\ \eqref{eq:blackcirclesquare4}) to the set of invariant states drastically simplifies
\begin{equation}
  \ket*{\blackcircleSA{\mathrm{d}}^{(4)}} \mapsto 
  \frac{\sqrt{q}}{q-1}\ket*{\combSA^{(4)}}-\frac{1}{q-1}\ket*{\circleSA^{(4)}}.
\end{equation}
Note that there are no free observable-dependent parameters upon specifying $\tr[\mathcal{O}_{\rm d}]=0$, and $\tr[\smash{\mathcal{O}_{\mathrm{d}}\mathcal{O}^{\dagger}_{\mathrm{d}}}]=q$.  

This implies that the averaged OTOC between $\mathrm{O}_{\rm d}$ and a (so far) generic $\mathrm{O}_{\gamma}$ takes the following form,
\begin{equation}
  \expval{{O}_{\mathrm{d}\nu}(x,t)}_{\rpc}=
  \frac{q}{q-1} - \frac{\sqrt{q}}{q-1} t_{\nu}(t-x+1,t+x,1),
\end{equation}
where $t_{\nu}(t-x+1,t+x,1)$ is a special case of the generalised partition function $t_{\nu}(x,y,k)$ with $0\le k\le y$, and an auxiliary function $z_{\nu}(x,y,k)$, given by the following diagrams,
\begin{equation}
  t_\nu(x,y,k)=q^{x+y}
  \mkern-8mu
  \begin{tikzpicture}[baseline={([yshift=-0.6ex]current bounding box.center)},scale=0.5]
    \prop{0}{0}{colPerm}{4}
    \foreach \x in {-1,1}{\prop{\x}{1}{colPerm}{4}}
    \foreach \x in {-2,0,2}{\prop{\x}{2}{colPerm}{4}}
    \foreach \x in {-3,-1,...,3}{\prop{\x}{3}{colPerm}{4}}
    \foreach \x in {-4,-2,0,2}{\prop{\x}{4}{colPerm}{4}}
    \foreach \x in {-5,-3,-1,1}{\prop{\x}{5}{colPerm}{4}}
    \foreach \x in {-6,-4,-2,0}{\prop{\x}{6}{colPerm}{4}}
    \foreach \x in {-5,-3,-1}{\prop{\x}{7}{colPerm}{4}}
    \foreach \x in {-4,-2}{\prop{\x}{8}{colPerm}{4}}
    \foreach \x in {-3}{\prop{\x}{9}{colPerm}{4}}
    \foreach \x in {0,...,6}{\square{-0.5-\x}{-0.5+\x}}
    \foreach \x in {7,...,9}{\circle{-13.5+\x}{-0.5+\x}}
    \foreach \x in {1,...,3}{\square{0.5+\x}{-0.5+\x}}
    \foreach \x in {4,...,7}{\circle{7.5-\x}{-0.5+\x}}
    \blacksquareD{0.5}{-0.5}
    \circle{-3.5}{9.5}
    \combD{-2.5}{9.5}
    \combD{-1.5}{8.5}
    \combD{-0.5}{7.5}
    \node at (0.8,-0.9) {\scalebox{0.9}{$\mathcal{O}_{\nu}$}};
    \draw[semithick,decorate,decoration={brace}] (-6.65,6.65) -- (-3.65,9.65) node[midway,xshift=-5pt,yshift=5pt,rotate=45] {$x$};
    \draw[semithick,decorate,decoration={brace}] (0.65,6.65) -- (3.65,3.65) node[midway,xshift=5pt,yshift=5pt,rotate=-45] {$y-k$};
    \draw[semithick,decorate,decoration={brace}] (-2.35,9.65) -- (-0.35,7.65) node[midway,xshift=5pt,yshift=5pt,rotate=-45] {$k$};
  \end{tikzpicture}\,,\qquad
  z_{\nu}(x,y,k)=q^{x+y}\mkern-18mu
  \begin{tikzpicture}[baseline={([yshift=-0.6ex]current bounding box.center)},scale=0.5]
    \prop{0}{0}{colPerm}{4}
    \foreach \x in {-1,1}{\prop{\x}{1}{colPerm}{4}}
    \foreach \x in {-2,0,2}{\prop{\x}{2}{colPerm}{4}}
    \foreach \x in {-3,-1,...,3}{\prop{\x}{3}{colPerm}{4}}
    \foreach \x in {-4,-2,0,2}{\prop{\x}{4}{colPerm}{4}}
    \foreach \x in {-5,-3,-1,1}{\prop{\x}{5}{colPerm}{4}}
    \foreach \x in {-6,-4,-2,0}{\prop{\x}{6}{colPerm}{4}}
    \foreach \x in {-5,-3}{\prop{\x}{7}{colPerm}{4}}
    \foreach \x in {-4}{\prop{\x}{8}{colPerm}{4}}
    \foreach \x in {0,...,6}{\square{-0.5-\x}{-0.5+\x}}
    \foreach \x in {7,...,9}{\circle{-13.5+\x}{-0.5+\x}}
    \foreach \x in {1,...,3}{\square{0.5+\x}{-0.5+\x}}
    \foreach \x in {4,...,7}{\circle{7.5-\x}{-0.5+\x}}
    \blacksquareD{0.5}{-0.5}
    \combD{-3.5}{8.5}
    \combD{-2.5}{7.5}
    \combD{-1.5}{6.5}
    \combD{-0.5}{6.5}
    \node at (0.8,-0.95) {\scalebox{0.9}{$\mathcal{O}_{\nu}$}};
    \draw[semithick,decorate,decoration={brace}] (-6.65,6.65) -- (-4.65,8.65) node[midway,xshift=-5pt,yshift=5pt,rotate=45] {$x-1$};
    \draw[semithick,decorate,decoration={brace}] (0.65,6.65) -- (3.65,3.65) node[midway,xshift=5pt,yshift=5pt,rotate=-45] {$y-k$};
    \draw[semithick,decorate,decoration={brace}] (-3.35,8.65) -- (-1.35,6.65) node[midway,xshift=5pt,yshift=5pt,rotate=-45] {$k$};
  \end{tikzpicture}\ .
\end{equation}
Using the relations~\eqref{eq:otocLocalRelations}, we can relate $t_{\nu}(x,y,k)$ and $r_{\nu}(x,y,k)$ for different values of $k$, $x$, and $y$ to each other as
\begin{equation}
  \begin{gathered}
  t_{\nu}(x,y,k)=\frac{\sqrt{q}}{1+q}
  \left[t_{\nu}(x,y-1,k-1)+z_{\nu}(x,y,k)\right],\qquad
  z_{\nu}(x,y,k)=\frac{\sqrt{q}}{1+q}
  \left[t_{\nu}(x-1,y,k)+z_{\nu}(x,y,k+1)\right],\\
  t_{\nu}(x,y,0)=1,\qquad
    t_{\nu}(0,y,k)=q^{\frac{k}{2}}
    \left[1+\delta_{k,y}(o_{\nu,2}-1)\right],\qquad
  z_{\nu}(x,y,y)=\sqrt{q} t_{\nu}(x-1,y,y),
  \end{gathered}
\end{equation}
where we introduced the $\mathcal{O}_{\nu}$-dependent parameter $o_{\nu,2}$ defined as
\begin{equation}
  o_{\nu,2} =\frac{1}{q} \sum_{x=1}^{q}\mel{x}{\mathcal{O}_{\nu}}{x}^2,\qquad
  0\le o_{\nu,2}\le 1.
\end{equation}
Note that $o_{\nu,2}=1$ whenever $\mathcal{O}_{\nu}$ is diagonal, and $o_{\nu,2}=0$ if $\mathcal{O}_{\nu}$ is orthogonal to \emph{all} diagonal observables (i.e.\ its diagonal elements are all $0$).

It is convenient to perform a change of variables
\begin{equation}
  u_k(x,y)=
  \frac{q+1}{q-1} \left(\frac{\sqrt{q}}{q+1}\right)^{-x-2y-2k}
  \left[q^{\frac{x}{2}}-t_{\nu}(y,k+x,x)\right],
\end{equation}
in terms of which the OTOC is expressed as
\begin{equation}
  \expval{{O}_{\mathrm{d}\nu}(x,t)}_{\rpc}=
  \left(\frac{\sqrt{q}}{q+1}\right)^{4t+2} u_{t+x-1}(1,t-x+1).
\end{equation}
Function $u_{k}(x,y)$ satisfies the relations
\begin{equation}
  \begin{gathered}
    u_{k}(x,y)=u_{k}(x-1,y)+(q+1)u_{0}(x+k,y-1)+\sum_{m=1}^k u_{m}(x+k-m,y-1),\\
    u_{k}(x,0)=\delta_{k,0}(1-\delta_{x,0})\frac{q+1}{q-1} \left(1-o_{\nu,2}\right)
  \left(q+1\right)^{x},\qquad
  u_{k}(0,y)=0,
  \end{gathered}
\end{equation}
which have the same form as Eqs.~\eqref{eq:singleRecRelation} and~\eqref{eq:singleRecRelationBCs} up to a redefinition of the parameters. In particular, the above relation is solved by Eq.~\eqref{eq:solution2pcorr} upon setting 
\begin{equation}
  \alpha\mapsto 1+q,\qquad\frac{q-1}{\sqrt{q}} \mapsto \frac{q+1}{q-1}(1-o_{\nu,2}).
\end{equation}

\section{Averaged OTOCs in Random Unitary Circuits}
\label{sec:OTOCsRUC}

Here we show that the averaged OTOC between local operators in random unitary circuit can be mapped to the solution of Eqs.~\eqref{eq:singleRecRelation} and~\eqref{eq:singleRecRelationBCs}. 

We begin by rewriting Eq.~\eqref{eq:OTOCRUCsys} as a single equation by substituting the second equation into the first. The result reads as  
\begin{equation}
\label{eq:asystemapp}
  a_k(x,y)=\frac{1}{1+q^2} a_{k}(x-1,y)+a_{0}(x,y-1) \left[\frac{q}{1+q^2}\right]^{k+1} +\frac{1}{q} \sum_{r=0}^{k-1} a_{k-r}(x,y-1) \left[\frac{q}{1+q^2}\right]^{r+2},
\end{equation}
while the relevant boundary conditions are given by  
\begin{equation}
\label{eq:asystemappBC}
    a_k(x,0)=\frac{1}{q^{|k-1|}},    \qquad
    a_k(k,y)=\frac{1}{q^{|y+k-1|}}. 
\end{equation}
This form of the recursion relation is the original one given in Ref.~\cite{bertini2020scrambling}. 

We now proceed introducing the new variable 
\be
\tilde z_k(x,y)=(q^2+1)^{x+y+k} \left(\frac{q^2+1}{q}\right)^{k+y} a_{k}(x+k,y).
\label{eq:ztilde}
\ee
Rewriting \eqref{eq:asystemapp} and \eqref{eq:asystemappBC} in terms of $\tilde z_k(x,y)$ we have 
\be
\label{eq:ztildesys}
 \tilde z_k(x,y)= \tilde z_{k}(x-1,y)+\tilde z_{0}(x+k,y-1) ({1+q^2}) + \sum_{r=0}^{k-1} \tilde z_{r}(x+k-r,y-1),
\ee
and 
\begin{equation}
\label{eq:ztildeBC}
\tilde z_k(x,0) =\frac{1}{q^{|k-1|}} (q^2+1)^x \left(\frac{(q^2+1)^2}{q}\right)^{k}, \qquad \tilde z_k(0,y) =\frac{1}{q^{|k+y-1|}} \left(\frac{(q^2+1)^2}{q}\right)^{k+y}. 
\end{equation}
Note that Eq.~\eqref{eq:ztildesys} coincides with Eq.~\eqref{eq:singleRecRelation} with the replacement 
\be
\alpha\mapsto (1+q^2). 
\ee
Our next observation is that Eq.~\eqref{eq:ztildesys} (but not the boundary conditions \eqref{eq:ztildeBC}) is solved by the following product of exponentials
\be
s_k(x,y)= \left(\frac{(q^2+1)^2}{q^2}\right)^{k+y} (q^2+1)^x\,. 
\ee
Therefore, also the function
\be
z_k(x,y) = \frac{q^2}{(q^2-1)^2} (q s_k(x,y)-\tilde z_k(x,y)), 
\ee
solves Eq.~\eqref{eq:ztildesys} by linearity. The boundary conditions for $z_k(x,y)$ are given by 
\begin{equation}
\label{eq:zBC}
z_{k}(x,0) =\delta_{k,0} \frac{q^2+1}{q^2-1} (q^2+1)^x, \qquad z_k(0,y) =0. 
\end{equation}
Noting that the value $z_{k}(0,0)$ does not affect $z_{k}(x,y)$ whenever $x$ or $y$ are non-zero, we have that Eq.~\eqref{eq:zBC} can be changed to 
\be
\label{eq:zBCmod}
z_{k}(x,0) =\delta_{k,0}(1-\delta_{x,0}) \frac{q^2+1}{q^2-1} (q^2+1)^x, \qquad z_k(0,y) =0. 
\ee
These conditions coincide with Eq.~\eqref{eq:singleRecRelationBCs} with the replacement $\alpha\mapsto1+q^2$ up to the multiplicative constant
\be
\frac{q^2+1}{q^2-1}\frac{\sqrt{q}}{q-1}\,.
\ee
Putting all together we have that $z_{k}(x,y)$ can be obtained from Eq.~\eqref{eq:solution2pcorr} with the replacement 
\begin{equation}
  \alpha\mapsto 1+q^2,\qquad\frac{q-1}{\sqrt{q}} \mapsto \frac{q^2+1}{q^2-1}.
\end{equation}
To conclude, we note that rewriting Eq.~\eqref{eq:OTOCRUCinitial} in terms of $z_{k}(x,y)$ we have  
\be
\expval{{O}_{\mu\nu}(x,t)}_{\ruc} \!=\! \left(\frac{q}{q^2+1}\right)^{4t+2} z_{t-x}(1,t+x),
\ee
Plugging the explicit form of $z_{k}(x,y)$, this expression gives precisely Eq.~\eqref{eq:exactRUCsRPCs}. 

\section{Purity: Flat initial state}
\label{app:purityFlat}

In this appendix we apply our large $q$ expansion method to characterise the entanglement of the uniform superposition of all computation-basis states, i.e.\
\begin{equation}
  \ket{\psi}=\frac{1}{\sqrt{q}} \sum_{i=0}^{q-1}\ket{i},\qquad
  \ket*{\inStateSA}=q \ket*{\flatSA},\qquad
  \ket*{\flatSA}:=\ket*{\pi_{15}}.
\end{equation}
In this case, since $\ket{\flatSA}\otimes\ket{\flatSA}$ is an eigenvector of $\mathcal{Q}^{(4)}$, the diagram simplifies and we get 
\begin{equation}
\begin{aligned}
  \left.\expval{P(t)}_{\rpc}\right|_{\text{flat}}&=
  q^{4t}
\begin{tikzpicture}[baseline={([yshift=-0.6ex]current bounding box.center)},scale=0.5]
    \prop{0}{0}{colPerm}{4}
    \foreach \x in {-1,1}{\prop{\x}{-1}{colPerm}{4}}
    \foreach \x in {-2,0,2}{\prop{\x}{-2}{colPerm}{4}}
    \foreach \x in {-3,-1,1,3}{\prop{\x}{-3}{colPerm}{4}}
    \foreach \x in {-4,-2,...,4}{\prop{\x}{-4}{colPerm}{4}}
    \foreach \x in {-5,...,-1}{\circle{\x+0.5}{\x+1.5}}
    \foreach \x in {-5,...,-1}{\squareD{-\x-0.5}{\x+1.5}}
    \foreach \x in {-5,-3,...,4}{\flatLD{\x+0.5}{-4.5}}
    \foreach \x in {-4,-2,...,4}{\flatRD{\x+0.5}{-4.5}}

\end{tikzpicture}
\\
&=q^{4t}
  \braket*{\circleSA}{\flatSA}^{2t} \braket*{\squareSA}{\flatSA}^{2t}=1,
  \end{aligned}
\end{equation}
which is expected, as the initial state is invariant under dynamics, and there
can be no entanglement growth. On the other hand, the solution~\eqref{eq:solPurity}
clearly does not recover this, as in this case
\begin{equation}
  \braket*{\combSA}{\inStateSA}\to
  q \braket*{\combSA}{\flatSA}=\frac{1}{\sqrt{q}}.
\end{equation}
This highlights a subtlety of our large-$q$ expansion in
Eq.~\eqref{eq:purityLocalRelations}. We are assuming that for a given vector $\ket{x}$
the largest contribution to
\begin{align}
  &\bra{x}\mathcal{Q}^{(4)}\ket*{\circleSA}\otimes\!\ket*{\squareSA}=\bra{x}\mathcal{Q}^{(4)}\ket*{\pi_6}\otimes\!\ket*{\pi_7}
  =\bra{x}\sum_{m=1}^{15}  C^{(4)}_{6,7,m} 
  \ket*{\pi_m}\otimes \ket*{\pi_m}, \\
 &C^{(4)}_{6,7,m} := \sum_{n=1}^{15}  [W_{\rm P}^{(n)}]_{mn} \braket{\pi_6}{\pi_n}\braket{\pi_7}{\pi_n}
\end{align}
comes from the terms with the largest value of $C^{(4)}_{6,7,m}$. This, however, is
neglecting the contribution of the overlap $\bra{x} (\ket*{\pi_m^{(4)}}\otimes
\ket*{\pi_m^{(4)}})$. For generic states (including random and computational basis states) this is not a problem because the terms that we neglected in~\eqref{eq:purityLocalRelations} do not have a large overlap with the appropriate states $\bra{x}$ that arise in the evaluation of the diagram. However, in the case of the flat initial state this is no longer the case, as $\braket*{\flatSA}{\circleSA}=\braket*{\flatSA}{\squareSA}=q^{-1}$, and $\braket*{\flatSA}{\combSA}=q^{-\frac{1}{2}}$, while $\braket*{\flatSA}{\flatSA}=1$, and therefore the contribution 
\begin{equation}
  C_{6,7,1}^{(4)}=\frac{1}{q^2}+\mathcal{O}(q^{-3})
\end{equation}
can no longer be neglected.

\twocolumngrid
\bibliography{./bibliography,./daniel-bib.bib}
\end{document}